\documentclass[twocolumn, prl, 10pt]{revtex4-1}
\usepackage{graphicx, subfigure, amsfonts,amssymb,amsmath, array, gensymb,fontenc,color, lipsum}
\usepackage{booktabs} 

\usepackage{booktabs}
\usepackage{multirow}
\begin{document}

\author{Michele~Pizzochero}
\email{mpizzochero@g.harvard.edu}
\affiliation{School of Engineering and Applied Sciences, Harvard University, Cambridge, Massachusetts 02138, United States}

\author{Efthimios~Kaxiras}
\affiliation{School of Engineering and Applied Sciences, Harvard University,  Cambridge, Massachusetts 02138, United States}

\title{Hydrogen Atoms on Zigzag Graphene Nanoribbons: \\ Chemistry and Magnetism Meet at the Edge}

\begin{abstract}
Although the unconventional $\pi$-magnetism at the zigzag edges of graphene holds promise for a wide array of applications, whether and to what degree it plays a role in their chemistry remains poorly understood. Here, we investigate the addition of a hydrogen atom --- the simplest yet the most experimentally relevant adsorbate --- to zigzag graphene nanoribbons (ZGNRs).  We show that the $\pi$-magnetism governs the chemistry of ZGNRs, giving rise to a site-dependent reactivity of the carbon atoms  and driving the hydrogenation process to the nanoribbon edges. Conversely, the chemisorbed hydrogen atom governs the $\pi$-magnetism of ZGNRs, acting as a spin-$\frac{1}{2}$ paramagnetic center in the otherwise antiferromagnetic ground state and spin-polarizing the charge carriers at the band extrema. Our findings establish a comprehensive picture of the peculiar interplay between chemistry and magnetism that emerges at the zigzag edges of graphene.
\end{abstract}

\maketitle


\paragraph{Introduction.}
The properties of graphene are largely shaped by the nature of the edges \cite{Fujii2013, Zhang2013, Jia2011}. Celebrated examples in this vein are graphene nanoribbons --- few-nanometer-wide strips of hexagonally bonded  carbon atoms --- where edge effects forge the electronic structure \cite{Yazyev2013, Wassmann2008, Wassmann2010}. In these systems, the lateral confinement of the charge carriers opens a width-dependent band gap \cite{Son2006a, Han2007, Pizzochero2020a}, which renders graphene nanoribbons optimally suited for logic applications in the ultimate limit of scalability \cite{Llinas2017, Jacobse2017, Zongping2020}. The rapid progress in on-surface synthesis \cite{Cai2010a} enabled the fabrication of atomically precise graphene nanoribbons that exhibit a broad spectrum of edge topologies \cite{Yano2020}, including armchair \cite{Talirz2017}, zigzag \cite{Ruffieux2016}, chiral \cite{Li2021}, and chevron \cite{Nguyen2017} edges, to mention but a few.

Within this variety of nanoribbons, zigzag graphene nanoribbons (ZGNRs) are unique owing to their intrinsic $\pi$-electron magnetism. \cite{Yazyev2010}. This is a result of a width-independent Stoner-like instability \cite{Fujita1996, Yazyev2010}, which promotes the emergence of magnetic moments  at the zigzag sites \cite{Li2019} that couple ferromagnetically along the edges and antiferromagnetically across the nanoribbon axis \cite{Fujita1996, Son2006a}. The resulting spin-$0$ magnetic ground state has been predicted to be preserved at room temperature \cite{Magda2014}, and can be extensively engineered via external \cite{Son2006} and built-in electric fields \cite{Kan2008}, lattice deformations \cite{Hu2012, Zhang2017}, charge doping \cite{Jung2010}, or incorporation of point defects \cite{Wimmer2008, Pizzochero2021a}. The combination of tunable magnetism, sizable band-gap, along with long spin lifetimes ensuing from weak spin-orbit and hyperfine interactions \cite{Avsar2020, Han2014}, establishes ZGNRs as prime candidates for spin-logic operations \cite{Yazyev2008, Wang2009}.

\begin{figure}[t]
    \centering
    \includegraphics[width=1\columnwidth]{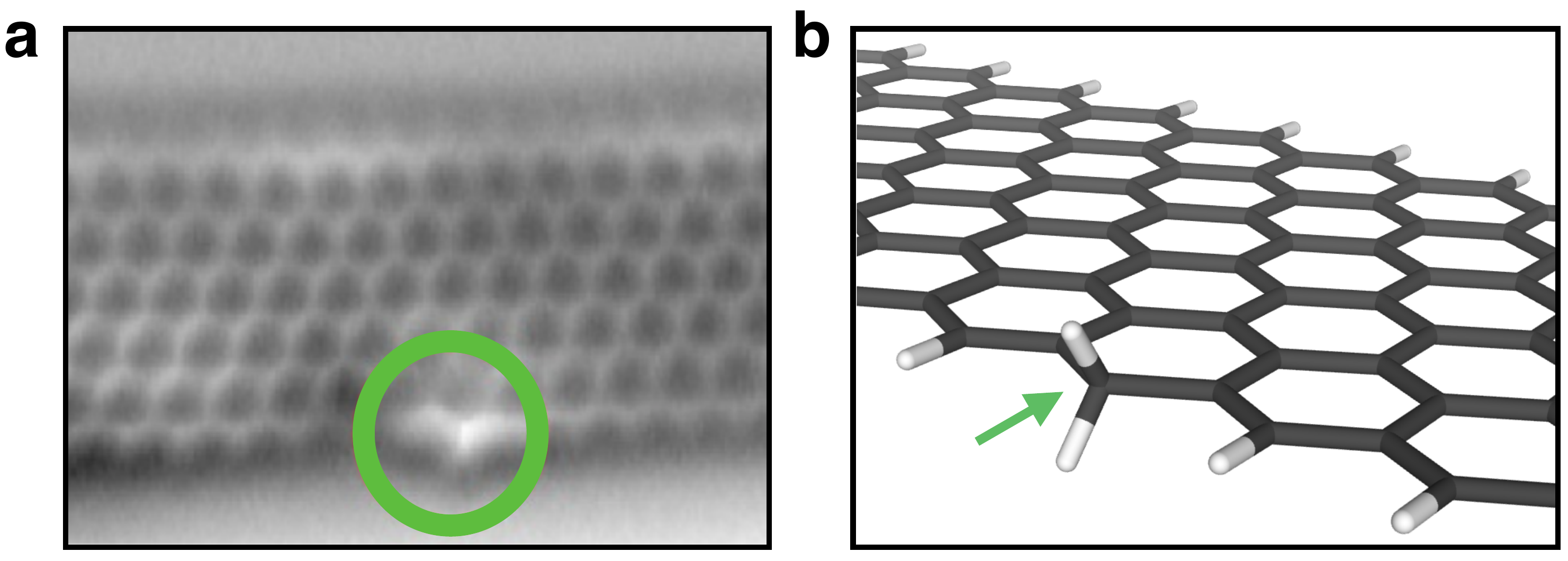}
    \caption{\textbf{Visualizing hydrogen adatoms in ZGNRs.} (a) Representative non-contact atomic force microscopy image of a hydrogen atom chemisorbed at the edge carbon atom of 6-ZGNRs, along with its (b) atomic structure obtained from first-principles calculations. Panel (a) is adapted with permission from Ref.\ \citenum{Ruffieux2016}. \label{Fig1}}
\end{figure}

Chemical functionalization is among the most effective routes to modulate the electronic structure of graphene and related nanostructures, with atomic hydrogen being the simplest yet the most widely used chemical species for this purpose \cite{Bonfanti2018a}. For instance, exposure of graphene to hydrogen can induce a reversible metal-to-insulator transition \cite{Elias2009, Balog2010}, enhance spin-orbit interactions \cite{Balakrishnan2013}, or introduce a controllable ferromagnetism \cite{Gonzales2016}.  Despite the constantly growing interest in the reactivity of graphene, the understanding of hydrogen adsorption on zigzag graphene nanoribbons is very limited \cite{Jiang2007, Campisi2020}. On the experimental side, ubiquitous hydrogen adatoms have been observed in atomically precise ZGNRs \cite{Ruffieux2016}, as shown in Figure \ref{Fig1}. These adatoms invariably form at the edges, thus hinting at a site-dependent chemical reactivity of the carbon atoms across the nanoribbon, the origin of which remains elusive. 

In this work, we combine first-principles and non-equilibrium Green's function calculations to elucidate the role of the $\pi$-magnetism of zigzag graphene nanoribbons on their chemistry. We consider the addition of a single hydrogen atom to ZGNRs and investigate the energetics of the adsorption process together with its impact on the electronic structure and quantum charge transport.  
On the one hand, the $\pi$-magnetism controls the reactivity of ZGNRs by introducing a site-dependence of the kinetics and thermodynamics of the chemisorption, which renders the hydrogenation a process mainly related to the edges of the nanoribbon.  Upon chemisorption, on the other hand, the hydrogen adatom controls the $\pi$-magnetism by acting as a paramagnetic spin-$\frac{1}{2}$ center in the otherwise antiferromagnetic ground state, and inducing a substantial spin-polarization of the charge carriers that confers spin-injection capabilities to ZGNRs.  Overall, our findings reveal that chemistry and magnetism are  intertwined at the zigzag edges of graphene. 

\smallskip
\paragraph{Methodology.} 
Our first-principles calculations are carried out within the spin-polarized density-functional theory (DFT) framework. We rely on the gradient-corrected approximation to the exchange-correlation functional devised by Perdew, Burke, and Ernzerhof (PBE) \cite{PBE} and account for the van der Waals interactions through the DFT-D3 dispersion correction method \cite{DFTD3}. Quantum electronic transport calculations are based on the Landauer formula within the non-equilibrium Green's function formalism. Our models consist of hydrogen-terminated ZGNRs that are six-carbon zigzag lines wide (6-ZGNRs). This is the sole width for which hydrogen adatoms have been experimentally characterized, as shown in Figure \ref{Fig1}, and that has been achieved in an atomically precise fashion through on-surface synthesis \cite{Ruffieux2016}. Further technical details of the methodology employed are provided in Supporting Note S1.

\begin{figure}[t!]
    \centering
    \includegraphics[width=1\columnwidth]{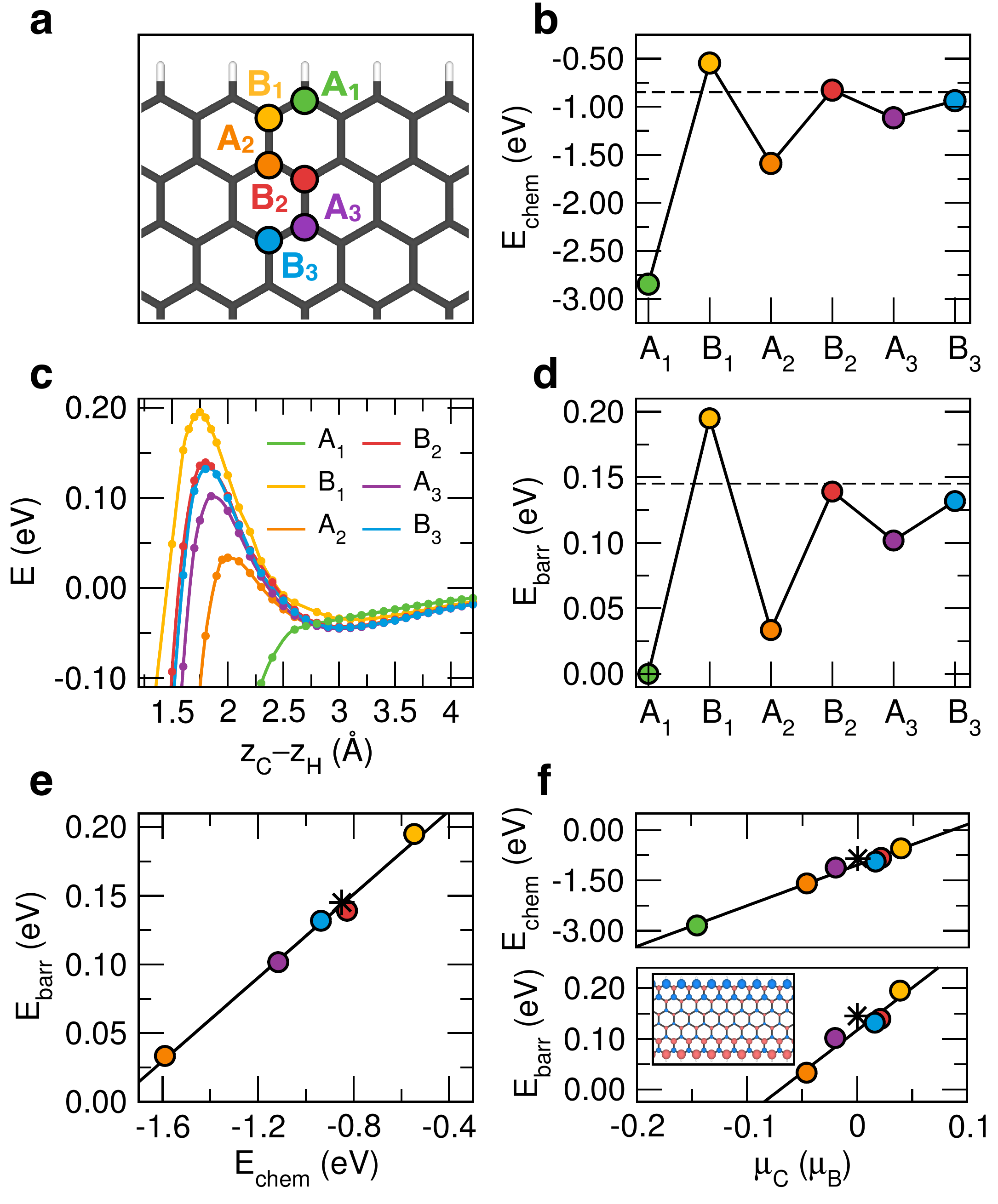}
    \caption{\textbf{Adsorption energetics of a hydrogen atom on ZGNRs.} (a) Illustration of the six symmetry-unique sites for hydrogen adsorption on 6-ZGNRs denoted as $A_n$ or $B_n$ ($n = 1, 2, 3$). (b) Chemisorption energies, $E\textsubscript{chem}$, of a hydrogen atom on ZGNRs; the horizontal dashed line indicates $E\textsubscript{chem}$ on graphene. (c) Energy profiles of the adsorption of a hydrogen atom on  ZGNRs.  (d) Energy barrier, $E\textsubscript{barr}$, of a hydrogen atom to chemisorb on ZGNRs; the horizontal dashed line indicates $E\textsubscript{barr}$ on graphene.  (e) Linear correlation of $E\textsubscript{barr}$ with $E\textsubscript{chem}$ as per Bell-Evans-Polanyi principle. (f) Linear correlation of $E\textsubscript{barr}$ and $E\textsubscript{chem}$ with the site-integrated magnetic moments, $\mu\textsubscript{C}$; in panels (e) and (f), circles indicate the values pertaining to hydrogen chemisorption on ZGNRs whereas the star marks the corresponding values for graphene. The inset in panel (f) shows the spin density of the pristine ZGNR, where the red (blue) isosurface represents the spin-majority (spin-minority) contribution.
\label{Fig2}}
\end{figure}

\smallskip
\paragraph{Energetics of the adsorption and diffusion of an H atom on ZGNRs.} 
We begin by systematically investigating the chemisorption of a hydrogen atom on ZGNRs.  We consider the six symmetry-unique chemisorption sites available in 6-ZGNRs, and refer to them as either  $A_n$ or $B_n$ ($n = 1, 2, 3$) to distinguish between the two sublattices, as illustrated in Figure \ref{Fig2}(a). The corresponding atomic structures upon the addition of the hydrogen atom are shown in Supporting Figure S1. For each site, we assess the thermodynamic stability of the adatom by determining the chemisorption energy as 
\begin{equation}
E\textsubscript{chem} = E\textsubscript{H+ZGNR} - (E\textsubscript{ZGNR} + E\textsubscript{H}), 
\end{equation}
where  $E\textsubscript{ZGNR}$ and  $E\textsubscript{H+ZGNR}$ are the total energies of the pristine and hydrogenated nanoribbon, respectively, and $E\textsubscript{H}$ is the energy of an isolated hydrogen atom. According to this expression, negative values of $E\textsubscript{chem}$ denote exothermic processes. In Figure \ref{Fig2}(b), we give the resulting chemisorption energies, which are found to span an interval as large as $\sim$$2$ eV, thus pointing to a very diverse reactivity of the carbon atoms across ZGNRs. The most stable site for the adatom to form is the edge carbon atom ($A_1$ site) where $E\textsubscript{chem}$ attains its minimum value of $-2.85$ eV, whereas the least stable site is the bridge carbon atom ($B_1$ site), for which $E\textsubscript{chem} = -0.55$ eV. The thermodynamic stability of the adatoms exhibits a marked sublattice dependence, with hydrogen chemisorption being more energetically favorable at the $A$ sublattice. Moving from the edge to the inner sites of ZGNRs, we observe an increase of $E\textsubscript{chem}$ at the $A$ sublattice and a decrease at the $B$ sublattice, with the chemisorption energy eventually approaching that of graphene, $-0.85$ eV.

We analyze in more detail the energetics of the adsorption process by determining the energy profile along the minimum energy path. This is accomplished by following the trajectory of a hydrogen atom nearing the carbon atom in the out-of-plane direction \cite{Bonfanti2018b, Hornekaer2006, Casolo2009}. Our results are given in Figure \ref{Fig2}(c). As the hydrogen atom proceeds towards the nanoribbon, two effects take place. First, we observe the formation of a physisorption state, except for the undercoordinated edge site $A_1$. This state arises when the incoming hydrogen atom is located $\sim$3.1 {\AA} above the carbon atom and features physisorption energies of $-35$ meV at the $B_1$ site and  $-45$ meV at the other sites. This latter value matches the one pertaining to graphene and agrees with the experimental estimate of $-40$ meV \cite{Ghio1980}. Second, we observe that the incoming hydrogen atom may encounter an energy barrier to bind to ZGNRs, the height of which depends on the specific site. Remarkably, chemisorption at the $A_1$ site is barrierless, indicating that the hydrogenation of the carbon atoms at the edge of the nanoribbon is both thermodynamically and kinetically favorable. This result provides firm theoretical ground to the preferential formation of the adatom at the edge carbon atom observed in recent experiments \cite{Ruffieux2016}; cf.\  Figure \ref{Fig1}(a).  The energy barriers, $E\textsubscript{barr}$, for the hydrogenation of ZGNRs are displayed in Figure \ref{Fig2}(d) and mirror the trend of the chemisorption energies shown in Figure \ref{Fig2}(b), that is, they increase (decrease) for chemisorption occurring at the $A$ ($B$) moving from the edge to the innermost sites of ZGNRs, eventually converging to that of graphene, $0.15$ eV.   In Figure \ref{Fig2}(e), we rationalize this correlation between the chemisorption energies and the energy barriers at the various sites of  ZGNRs through the Bell-Evans-Polanyi principle \cite{Bell1936, Evans1936} according to which, for a set of thermally activated chemical reactions of the same family, these two quantities are linearly related as 
\begin{equation}
E\textsubscript{barr} = E_0 + \alpha E\textsubscript{chem},
\end{equation}
where $E_0 = 0.27$ eV and $\alpha = 0.15$ are the intercept and slope of the line describing the energetics of this family of reactions, respectively. Importantly, even the point associated with the energetics of the hydrogen chemisorption on graphene lies on this line, suggesting that such a  correlation is inherent to the addition of a hydrogen atom to $sp^2$-hybridized carbon nanostructures, regardless of their global electronic properties.

\begin{figure}[t!]
    \centering
    \includegraphics[width=1\columnwidth]{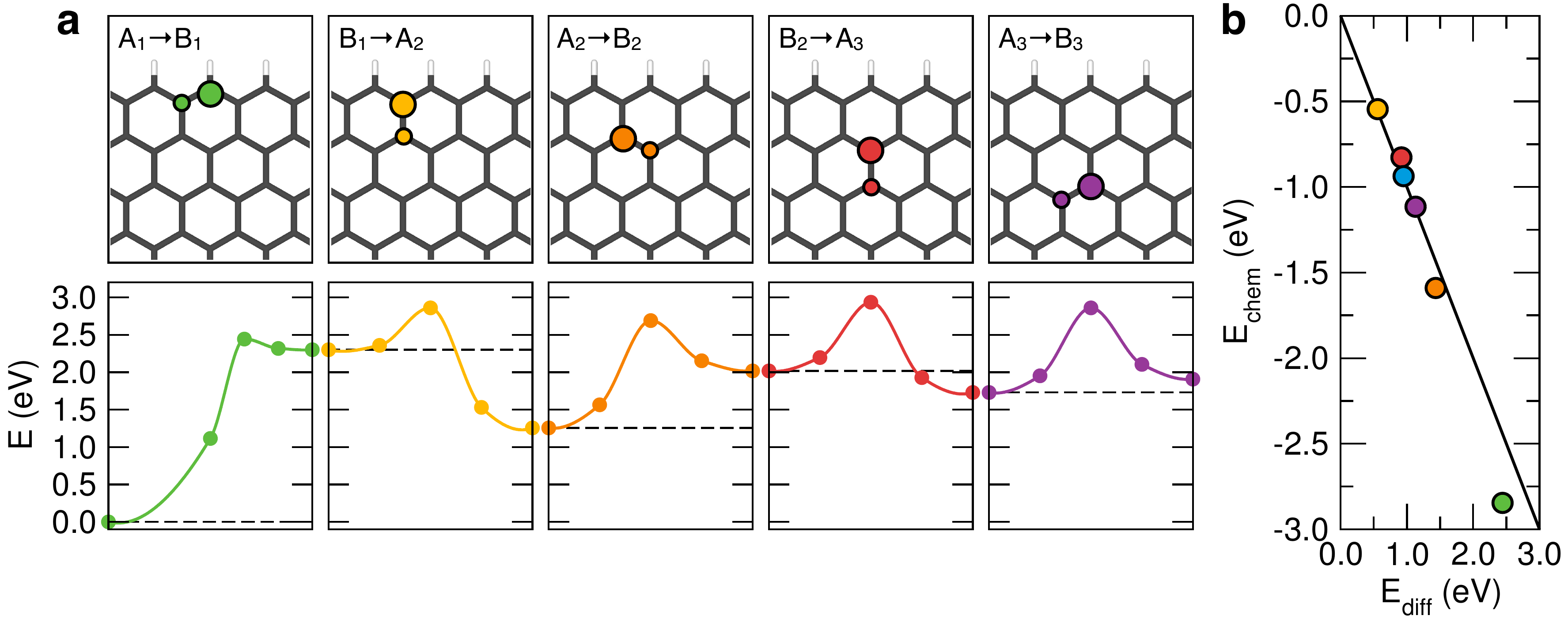}
    \caption{\textbf{Diffusion energetics of a hydrogen adatom on ZGNRs.} (a) Top panel: schematic illustration of the adatom diffusion pathways, with large and small circle indicating the initial and final sites at which the hydrogen adatom forms, respectively. Bottom panel: corresponding energy profiles of the adatom diffusion; lattice positions of ZGNRs are labeled according to Figure \ref{Fig2}(a). (b) Correlation between the diffusion barrier, $E\textsubscript{diff}$, and the chemisorption energy, $E\textsubscript{chem}$; black solid line indicates $E\textsubscript{diff} = E\textsubscript{chem}$.  Notice that the diffusion along the $B_3 \rightarrow A_3$ pathway (not shown) is symmetry-equivalent to the reverse of the $A_3 \rightarrow B_3$ pathway (blue color).
 \label{Fig3}}
\end{figure}

Our findings reveal that the chemisorption of a hydrogen atom on ZGNRs preferentially occurs at the $A$ sublattice, namely the one that comprises the edge carbon atoms. As shown in the inset of Figure \ref{Fig2}(f), this is the sublattice at which the spin density of ZGNRs primarily localizes \cite{Bonfanti2011}, signaling that the chemistry of zigzag graphene nanoribbons is modulated by its intrinsic magnetism. This effect is set on a quantitative basis in Figure \ref{Fig2}(f), where we show that both the $E\textsubscript{chem}$ and $E\textsubscript{barr}$ correlate linearly with the site-integrated magnetic moment, $\mu \textsubscript{C}$, namely, the net magnetic moment residing at a given carbon atom prior to chemisorption \cite{Casolo2009}. Hence, hydrogenation is facile at the sites which possess the most pronounced radical character {as a consequence of the coupling between the unpaired electron of the incoming hydrogen atom and that of the carbon atom}.  To ascertain whether or not structural effects play a role in the diverse reactivity of the carbon atoms in ZGNRs, we determine the reorganization energy,  $E\textsubscript{reorg}$ \cite{Bonfanti2011, Pizzochero2015}. This quantity accounts for the energy cost of the structural distortion in ZGNRs enforced by the adatom formation and is defined as 
\begin{equation}
E\textsubscript{reorg} = E_{{\textnormal{ZGNR}}}^* -  E\textsubscript{ZGNR}
\end{equation}
with $E\textsubscript{ZGNR}$ being the energy of the pristine ZGNR and $E\textsuperscript{*}\textsubscript{ZGNRs}$ the energy of the distorted configuration that the nanoribbon assumes upon the addition of a hydrogen atom. We obtain $E\textsubscript{reorg} = 1.0$ eV, irrespective of the chemisorption site; see Supporting Table S2. This site-independent value matches that of graphene, thereby demonstrating that structural effects are irrelevant to ruling the chemistry of ZGNRs, and the sublattice-dependent regioselectivity of the hydrogenation process is dominated by electronic effects solely \cite{Bonfanti2011}.

We next explore the mobility of the chemisorbed hydrogen atom on ZGNRs. We introduce the adatom at the thermodynamically and kinetically favorable $A_1$ site (the edge carbon atom) and cover all symmetry-unique sites upon successive hopping of the extra hydrogen atom to the nearest-neighbor carbon atom. The resulting diffusion pathways along with their corresponding energy profiles are shown in Figure \ref{Fig3}(a). The diffusion barrier, $E\textsubscript{diff}$, is the largest along the $A_1 \rightarrow B_1$ pathway ($E\textsubscript{diff} = 2.44$ eV) and the smallest along the $B_1 \rightarrow A_2$ one ($E\textsubscript{diff} = 0.56$ eV).  As shown in Figure \ref{Fig3}(b), at each site $E\textsubscript{diff}$ is approximately equal to the chemisorption energy (and governed by the same localization effects; see Supporting Figure S2), as, for the adatom to diffuse, the covalent bond between the carbon atom the extra hydrogen atom has to be broken. As a consequence, the diffusion process is inactive at room temperature and above, and, similarly to graphene \cite{Hornekaer2006}, adatoms on ZGNRs are more likely to desorb rather than diffuse.

\begin{figure}[t!]
    \centering
    \includegraphics[width=1\columnwidth]{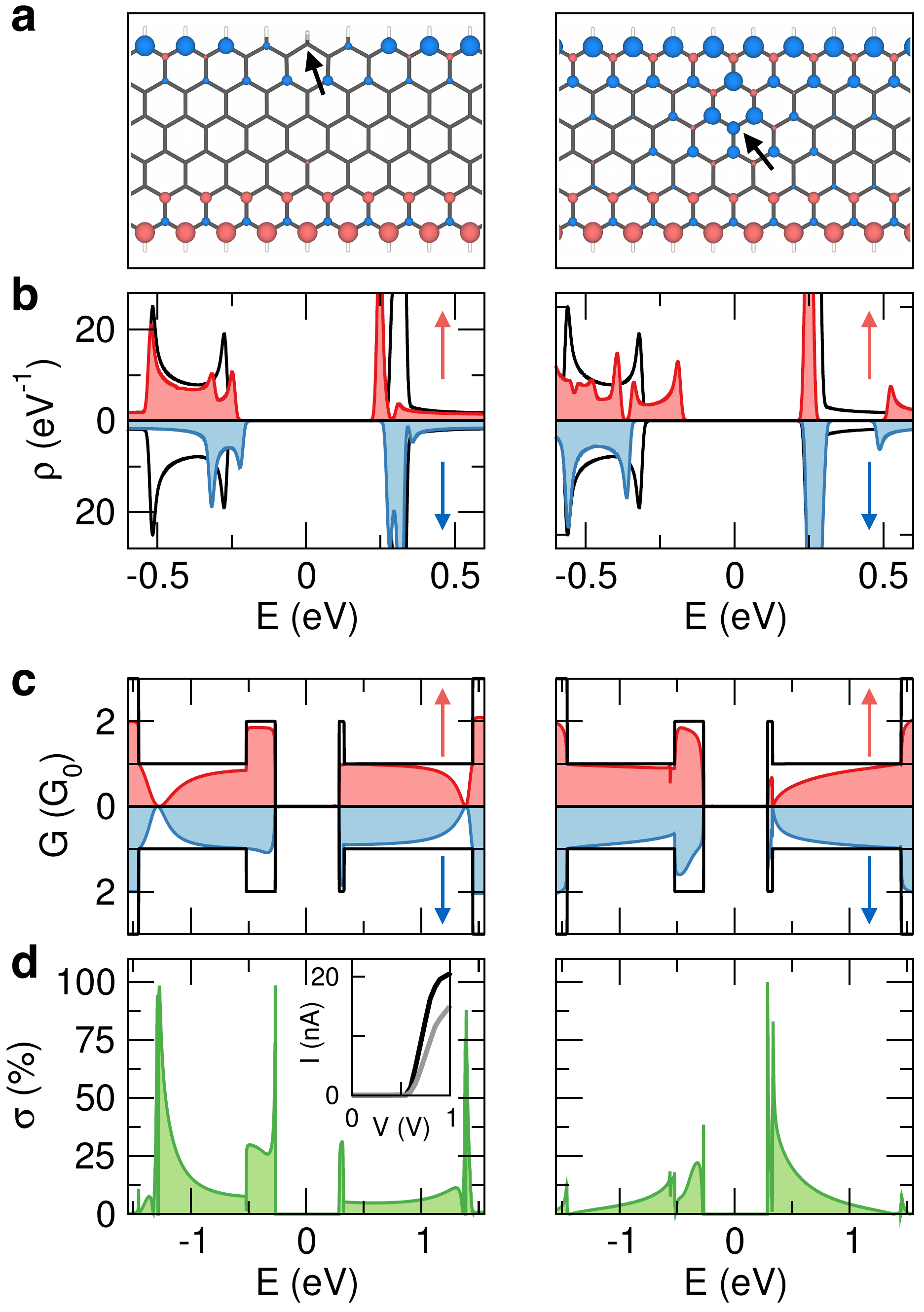}
    \caption{\textbf{Effect of the hydrogen adatom on the magnetism and charge transport in ZGNRs.} (a) {Spin density of ZGNRs upon hydrogen chemisorption at the edge ($A_1$ site) or the innermost ($B_3$ site) carbon atom, as marked by the black arrow; isosurfaces are set to $\pm 0.006$ $e$/{\AA}$^3$}. (b) Spin-resolved density of states, $\rho$, and (d) spin-resolved conductance, $G$, of ZGNRs hosting an adatom at the $A_1$ (left panel) and $B_3$ sites (right panel); also shown as a black line is the the corresponding values for a pristine ZGNR. (d) Spin-polarization of the conductance, $\sigma$, of ZGNRs hosting the adatom at the $A_1$ (left panel) and $B_3$ sites (right panel). The inset shows the $I$-$V$ characteristics of ZGNRs with (grey line) and without (black line) the adatom at the edge carbon atom ($A_1$ site). In panels (a-c), the red (blue) color indicates the spin-majority (spin-minority) contribution. \label{Fig4}} 
\end{figure}

\smallskip
\paragraph{Effect of an H adatom on the magnetism and quantum electronic transport in ZGNRs.} 
We examine the influence of the adatom on the electronic structure of ZGNRs. Irrespective of the site at which chemisorption occurs, the adatom acts a spin-$\frac{1}{2}$ center in the otherwise spin-0 antiferromagnetic ground state of ZGNRs. The chemisorption of a hydrogen atom requires a $\pi$-bond to break, removing one $p_z$ orbital from the $\pi$-electron network and releasing an unpaired electron on ZGNRs, which causes a local magnetic moment of 1 $\mu$\textsubscript{B} \cite{Casolo2009}. This result is in accordance with Lieb's theorem for the repulsive Hubbard model at a half filling \cite{Lieb1989} --- a model that has been widely used to describe the electronic properties of ZGNRs \cite{Fujita1996, Yazyev2010} --- which states that the spin ground state, $S$, of a bipartite lattice is imposed by the sublattice imbalance, 
\begin{equation}
S = \frac{1}{2} \left|n\textsubscript{A} - n\textsubscript{B}\right|, 
\end{equation}
with $n\textsubscript{A}$ and $n\textsubscript{B}$ being the number of $p_z$ orbitals at the $A$ and $B$ sublattices, respectively. 

In Figure \ref{Fig4}(a), we display the spin density upon hydrogenation at the two most representative sites, i.e., the edge ($A_1$ site) and innermost ($B_3$ site) carbon atoms; spin densities for other adatom configurations are shown in Supporting Figure S3. The spin density is found to localize mainly on the spin majority sublattice and to decay away from the extra hydrogen atom. {Depending on whether the chemisorption occurs at the edge or the innermost carbon atom, the magnetic moment is either {locally suppressed} ($A_1$ site) or gives rise to the characteristic $\sqrt{3} \times \sqrt{3} R 30\textsuperscript{o}$ pattern ($B_3$ site), respectively.} This latter case is analogous to that of graphene hosting point defects that remove a $p_z$ orbital from the $\pi$-electron cloud \cite{Casolo2009, Bonfanti2018a, Yazyev2010}, e.g., monovalent adatoms or carbon vacancies \cite{Yazyev2007}.  Despite these differences, comparable changes are observed in the density of states shown in Figure \ref{Fig4}(b). Independent of the chemisorption site, the extra hydrogen atom introduces localized states in the vicinity of the valence band maximum \cite{Buchs2007} and leads to a spin-splitting of the conduction states of ZGNRs; see Supporting Figure S4 for additional results.

Finally, we investigate the impact of the adatom on the charge transport across the nanoribbon, a critical issue to the performance of ZGNR-based nanoscale devices. In Figure \ref{Fig4}(c), we show the zero-bias conductance spectrum upon the chemisorption of a hydrogen atom at the edge ($A_1$ site) and innermost ($B_3$ site) carbon atoms; conductance spectra for other adatom configurations are given in Supporting Figure S5. The adatom leads to a depletion of the otherwise quantized conductance spectrum of ZGNRs, implying a substantial disruption of the charge transport. Similarly to previous works \cite{Pizzochero2020, Pizzochero2021a}, we quantify this effect through $\tau$, a dimensionless descriptor that evaluates the fraction of the ballistic conductance that is preserved in the vicinity ($\delta E = 0.1$ eV) of the valence band maximum (VBM) and conduction band minimum (CBM) of ZGNRs upon hydrogen chemisorption,
\begin{equation}
\tau = \frac{\int_{\textnormal{VBM}-\delta E}^{\textnormal{CBM} +\delta E} G\textsubscript{d}(E) dE}{\int_{\textnormal{VBM}-\delta E}^{\textnormal{CBM} +\delta E} G\textsubscript{p}(E) dE}  \%,
\end{equation}
where $G\textsubscript{d}(E)$ and $G\textsubscript{p}(E)$ are the zero-bias conductance spectrum of hydrogenated and pristine nanoribbon, respectively. We obtain $\tau$ of $72\%$ and $48\%$ upon chemisorption at the edge ($A_1$ site) and innermost ($B_3$ site) carbon atoms of ZGNRs, respectively. This indicates that the detrimental role of the adatom on the charge transport is weaker when hydrogen chemisorption occurs at the edges of the nanoribbon; see Supporting Figure S6. The deleterious effect of the adatom on the charge transport is reflected in the $I$-$V$ characteristics given in the inset of Figure \ref{Fig4}(c). The intensity current that develops when the applied bias voltage exceeds in magnitude the band-gap width of ZGNRs is significantly reduced upon hydrogenation, up to $25 \%$ at a typical $|V| = 1$ V. We notice that the formation of the adatom at the edge has a less severe impact on the electronic transport as compared to other common sources of structural disorder that emerge in the on-surface synthesis of nanoribbons, including "bite defects" (i.e., clusters of carbon vacancies located at the edges) in ZGNRs  \cite{Pizzochero2020, Pizzochero2021a}, which yield a decrease of the intensity current in the $I$-$V$ curve of over one order of magnitude and $\tau = 54\%$ \cite{Pizzochero2021a}.

The chemisorption of the hydrogen atom and the accompanying spin-$\frac{1}{2}$ magnetism introduces an energy-dependent spin-polarization of the charge carriers, $\sigma$, defined as\cite{Pizzochero2021a, Pizzochero2021b}
\begin{equation}
\sigma(E) = \left |\frac{G\textsubscript{d}{^\uparrow}(E) - G\textsubscript{d}{^\downarrow}(E)}{G\textsubscript{d}{^\uparrow}(E) + G\textsubscript{d}{^\downarrow}(E)}  \right | \%,
\end{equation}
with $G\textsubscript{d}{^\uparrow}(E)$ and $G\textsubscript{d}{^\downarrow}(E)$ being the zero-bias conductance spectrum of the spin-majority and spin-minority channels of the ZGNR hosting the adatom, respectively. In Figure \ref{Fig4}(d), we show the spin-polarization of the ballistic conductance, which is found to attain a value of unity at, e.g., the valence band maximum of ZGNRs upon the experimentally observed hydrogenation of the edge carbon atom ($A_1$ site). This result indicates that the formation of adatoms, even in the dilute regime, is remarkably effective in promoting a spin-polarized electronic transport across ZGNRs. We thus envisage moderately hydrogenated ZGNRs as a potential platform for carbon-based spin injection.

\smallskip
\paragraph{Summary and conclusions.}
We have unveiled a mutual interplay between chemistry and magnetism at the zigzag edges of graphene.  By considering the addition of a single hydrogen atom  --- the simplest yet the most experimentally relevant adsorbate --- to zigzag graphene nanoribbons (ZGNRs), we have shown that the diverse reactivity of the carbon atoms is dictated by the non-uniform spatial distribution of the intrinsic magnetic moments across ZGNRs. 
As a result, the chemisorption preferentially occurs at the sites where the spin density primarily localizes, with the edge carbon atoms being the most prone to hydrogenation, consistent with recent experiments \cite{Ruffieux2016}.
Upon chemisorption, the extra hydrogen atom is immobile and causes a substantial disruption of the electronic transport across the nanoribbon. The  adatom formation induces a paramagnetic response in the otherwise antiferromagnetic ground state of ZGNRs and gives rise to a sizable spin-polarization of the charge carriers, hence making mildly hydrogenated ZGNRs a highly promising candidate for spin-injection applications in all-carbon logic spintronics. 
To conclude, our findings are expected to be general to the adsorption process of other common monovalent species (e.g., OH$\cdot$, CH$_3$$\cdot$, Cl$\cdot$, F$\cdot$) and nanoribbon widths, thus offering a comprehensive understanding of the distinct chemistry of the magnetic zigzag edges of graphene.

\smallskip
\paragraph{Acknowledgments.}  
M.~P.~ is supported by the Swiss National Science Foundation (SNSF) through the Early Postdoc.Mobility program (Grant No.~P2ELP2-191706) and the NSF DMREF (Grant No. 1922165).
.

\bibliography{References.bib}

\begin{thebibliography}{55}%
\makeatletter
\providecommand \@ifxundefined [1]{%
 \@ifx{#1\undefined}
}%
\providecommand \@ifnum [1]{%
 \ifnum #1\expandafter \@firstoftwo
 \else \expandafter \@secondoftwo
 \fi
}%
\providecommand \@ifx [1]{%
 \ifx #1\expandafter \@firstoftwo
 \else \expandafter \@secondoftwo
 \fi
}%
\providecommand \natexlab [1]{#1}%
\providecommand \enquote  [1]{``#1''}%
\providecommand \bibnamefont  [1]{#1}%
\providecommand \bibfnamefont [1]{#1}%
\providecommand \citenamefont [1]{#1}%
\providecommand \href@noop [0]{\@secondoftwo}%
\providecommand \href [0]{\begingroup \@sanitize@url \@href}%
\providecommand \@href[1]{\@@startlink{#1}\@@href}%
\providecommand \@@href[1]{\endgroup#1\@@endlink}%
\providecommand \@sanitize@url [0]{\catcode `\\12\catcode `\$12\catcode
  `\&12\catcode `\#12\catcode `\^12\catcode `\_12\catcode `\%12\relax}%
\providecommand \@@startlink[1]{}%
\providecommand \@@endlink[0]{}%
\providecommand \url  [0]{\begingroup\@sanitize@url \@url }%
\providecommand \@url [1]{\endgroup\@href {#1}{\urlprefix }}%
\providecommand \urlprefix  [0]{URL }%
\providecommand \Eprint [0]{\href }%
\providecommand \doibase [0]{http://dx.doi.org/}%
\providecommand \selectlanguage [0]{\@gobble}%
\providecommand \bibinfo  [0]{\@secondoftwo}%
\providecommand \bibfield  [0]{\@secondoftwo}%
\providecommand \translation [1]{[#1]}%
\providecommand \BibitemOpen [0]{}%
\providecommand \bibitemStop [0]{}%
\providecommand \bibitemNoStop [0]{.\EOS\space}%
\providecommand \EOS [0]{\spacefactor3000\relax}%
\providecommand \BibitemShut  [1]{\csname bibitem#1\endcsname}%
\let\auto@bib@innerbib\@empty
\bibitem [{\citenamefont {Fujii}\ and\ \citenamefont
  {Enoki}(2013)}]{Fujii2013}%
  \BibitemOpen
  \bibfield  {author} {\bibinfo {author} {\bibfnamefont {S.}~\bibnamefont
  {Fujii}}\ and\ \bibinfo {author} {\bibfnamefont {T.}~\bibnamefont {Enoki}},\
  }\href@noop {} {\bibfield  {journal} {\bibinfo  {journal} {Accounts of
  Chemical Research}\ }\textbf {\bibinfo {volume} {46}},\ \bibinfo {pages}
  {2202} (\bibinfo {year} {2013})}\BibitemShut {NoStop}%
\bibitem [{\citenamefont {Zhang}\ \emph {et~al.}(2013)\citenamefont {Zhang},
  \citenamefont {Xin},\ and\ \citenamefont {Ding}}]{Zhang2013}%
  \BibitemOpen
  \bibfield  {author} {\bibinfo {author} {\bibfnamefont {X.}~\bibnamefont
  {Zhang}}, \bibinfo {author} {\bibfnamefont {J.}~\bibnamefont {Xin}}, \ and\
  \bibinfo {author} {\bibfnamefont {F.}~\bibnamefont {Ding}},\ }\href@noop {}
  {\bibfield  {journal} {\bibinfo  {journal} {Nanoscale}\ }\textbf {\bibinfo
  {volume} {5}},\ \bibinfo {pages} {2556} (\bibinfo {year} {2013})}\BibitemShut
  {NoStop}%
\bibitem [{\citenamefont {Jia}\ \emph {et~al.}(2011)\citenamefont {Jia},
  \citenamefont {Campos-Delgado}, \citenamefont {Terrones}, \citenamefont
  {Meunier},\ and\ \citenamefont {Dresselhaus}}]{Jia2011}%
  \BibitemOpen
  \bibfield  {author} {\bibinfo {author} {\bibfnamefont {X.}~\bibnamefont
  {Jia}}, \bibinfo {author} {\bibfnamefont {J.}~\bibnamefont {Campos-Delgado}},
  \bibinfo {author} {\bibfnamefont {M.}~\bibnamefont {Terrones}}, \bibinfo
  {author} {\bibfnamefont {V.}~\bibnamefont {Meunier}}, \ and\ \bibinfo
  {author} {\bibfnamefont {M.~S.}\ \bibnamefont {Dresselhaus}},\ }\href@noop {}
  {\bibfield  {journal} {\bibinfo  {journal} {Nanoscale}\ }\textbf {\bibinfo
  {volume} {3}},\ \bibinfo {pages} {86} (\bibinfo {year} {2011})}\BibitemShut
  {NoStop}%
\bibitem [{\citenamefont {Yazyev}(2013)}]{Yazyev2013}%
  \BibitemOpen
  \bibfield  {author} {\bibinfo {author} {\bibfnamefont {O.~V.}\ \bibnamefont
  {Yazyev}},\ }\href@noop {} {\bibfield  {journal} {\bibinfo  {journal}
  {Accounts of Chemical Research}\ }\textbf {\bibinfo {volume} {46}},\ \bibinfo
  {pages} {2319} (\bibinfo {year} {2013})}\BibitemShut {NoStop}%
\bibitem [{\citenamefont {Wassmann}\ \emph {et~al.}(2008)\citenamefont
  {Wassmann}, \citenamefont {Seitsonen}, \citenamefont {Saitta}, \citenamefont
  {Lazzeri},\ and\ \citenamefont {Mauri}}]{Wassmann2008}%
  \BibitemOpen
  \bibfield  {author} {\bibinfo {author} {\bibfnamefont {T.}~\bibnamefont
  {Wassmann}}, \bibinfo {author} {\bibfnamefont {A.~P.}\ \bibnamefont
  {Seitsonen}}, \bibinfo {author} {\bibfnamefont {A.~M.}\ \bibnamefont
  {Saitta}}, \bibinfo {author} {\bibfnamefont {M.}~\bibnamefont {Lazzeri}}, \
  and\ \bibinfo {author} {\bibfnamefont {F.}~\bibnamefont {Mauri}},\ }\href
  {\doibase 10.1103/PhysRevLett.101.096402} {\bibfield  {journal} {\bibinfo
  {journal} {Physical Review Letters}\ }\textbf {\bibinfo {volume} {101}},\
  \bibinfo {pages} {096402} (\bibinfo {year} {2008})}\BibitemShut {NoStop}%
\bibitem [{\citenamefont {Wassmann}\ \emph {et~al.}(2010)\citenamefont
  {Wassmann}, \citenamefont {Seitsonen}, \citenamefont {Saitta}, \citenamefont
  {Lazzeri},\ and\ \citenamefont {Mauri}}]{Wassmann2010}%
  \BibitemOpen
  \bibfield  {author} {\bibinfo {author} {\bibfnamefont {T.}~\bibnamefont
  {Wassmann}}, \bibinfo {author} {\bibfnamefont {A.~P.}\ \bibnamefont
  {Seitsonen}}, \bibinfo {author} {\bibfnamefont {A.~M.}\ \bibnamefont
  {Saitta}}, \bibinfo {author} {\bibfnamefont {M.}~\bibnamefont {Lazzeri}}, \
  and\ \bibinfo {author} {\bibfnamefont {F.}~\bibnamefont {Mauri}},\ }\href
  {\doibase 10.1021/ja909234y} {\bibfield  {journal} {\bibinfo  {journal}
  {Journal of the American Chemical Society}\ }\textbf {\bibinfo {volume}
  {132}},\ \bibinfo {pages} {3440} (\bibinfo {year} {2010})}\BibitemShut
  {NoStop}%
\bibitem [{\citenamefont {Son}\ \emph {et~al.}(2006{\natexlab{a}})\citenamefont
  {Son}, \citenamefont {Cohen},\ and\ \citenamefont {Louie}}]{Son2006a}%
  \BibitemOpen
  \bibfield  {author} {\bibinfo {author} {\bibfnamefont {Y.-W.}\ \bibnamefont
  {Son}}, \bibinfo {author} {\bibfnamefont {M.~L.}\ \bibnamefont {Cohen}}, \
  and\ \bibinfo {author} {\bibfnamefont {S.~G.}\ \bibnamefont {Louie}},\
  }\href@noop {} {\bibfield  {journal} {\bibinfo  {journal} {Physical Review
  Letters}\ }\textbf {\bibinfo {volume} {97}},\ \bibinfo {pages} {216803}
  (\bibinfo {year} {2006}{\natexlab{a}})}\BibitemShut {NoStop}%
\bibitem [{\citenamefont {Han}\ \emph {et~al.}(2007)\citenamefont {Han},
  \citenamefont {\"Ozyilmaz}, \citenamefont {Zhang},\ and\ \citenamefont
  {Kim}}]{Han2007}%
  \BibitemOpen
  \bibfield  {author} {\bibinfo {author} {\bibfnamefont {M.~Y.}\ \bibnamefont
  {Han}}, \bibinfo {author} {\bibfnamefont {B.}~\bibnamefont {\"Ozyilmaz}},
  \bibinfo {author} {\bibfnamefont {Y.}~\bibnamefont {Zhang}}, \ and\ \bibinfo
  {author} {\bibfnamefont {P.}~\bibnamefont {Kim}},\ }\href {\doibase
  10.1103/PhysRevLett.98.206805} {\bibfield  {journal} {\bibinfo  {journal}
  {Physical Review Letters}\ }\textbf {\bibinfo {volume} {98}},\ \bibinfo
  {pages} {206805} (\bibinfo {year} {2007})}\BibitemShut {NoStop}%
\bibitem [{\citenamefont {\v{C}er\c{n}evi\v{c}s}\ \emph
  {et~al.}(2020)\citenamefont {\v{C}er\c{n}evi\v{c}s}, \citenamefont {Yazyev},\
  and\ \citenamefont {Pizzochero}}]{Pizzochero2020a}%
  \BibitemOpen
  \bibfield  {author} {\bibinfo {author} {\bibfnamefont {K.}~\bibnamefont
  {\v{C}er\c{n}evi\v{c}s}}, \bibinfo {author} {\bibfnamefont {O.~V.}\
  \bibnamefont {Yazyev}}, \ and\ \bibinfo {author} {\bibfnamefont
  {M.}~\bibnamefont {Pizzochero}},\ }\href@noop {} {\bibfield  {journal}
  {\bibinfo  {journal} {Physical Review B}\ }\textbf {\bibinfo {volume}
  {102}},\ \bibinfo {pages} {201406} (\bibinfo {year} {2020})}\BibitemShut
  {NoStop}%
\bibitem [{\citenamefont {Llinas}\ \emph {et~al.}(2017)\citenamefont {Llinas},
  \citenamefont {Fairbrother}, \citenamefont {Borin~Barin}, \citenamefont
  {Shi}, \citenamefont {Lee}, \citenamefont {Wu}, \citenamefont {Yong~Choi},
  \citenamefont {Braganza}, \citenamefont {Lear}, \citenamefont {Kau},
  \citenamefont {Choi}, \citenamefont {Chen}, \citenamefont {Pedramrazi},
  \citenamefont {Dumslaff}, \citenamefont {Narita}, \citenamefont {Feng},
  \citenamefont {M{\"u}llen}, \citenamefont {Fischer}, \citenamefont {Zettl},
  \citenamefont {Ruffieux}, \citenamefont {Yablonovitch}, \citenamefont
  {Crommie}, \citenamefont {Fasel},\ and\ \citenamefont {Bokor}}]{Llinas2017}%
  \BibitemOpen
  \bibfield  {author} {\bibinfo {author} {\bibfnamefont {J.~P.}\ \bibnamefont
  {Llinas}}, \bibinfo {author} {\bibfnamefont {A.}~\bibnamefont {Fairbrother}},
  \bibinfo {author} {\bibfnamefont {G.}~\bibnamefont {Borin~Barin}}, \bibinfo
  {author} {\bibfnamefont {W.}~\bibnamefont {Shi}}, \bibinfo {author}
  {\bibfnamefont {K.}~\bibnamefont {Lee}}, \bibinfo {author} {\bibfnamefont
  {S.}~\bibnamefont {Wu}}, \bibinfo {author} {\bibfnamefont {B.}~\bibnamefont
  {Yong~Choi}}, \bibinfo {author} {\bibfnamefont {R.}~\bibnamefont {Braganza}},
  \bibinfo {author} {\bibfnamefont {J.}~\bibnamefont {Lear}}, \bibinfo {author}
  {\bibfnamefont {N.}~\bibnamefont {Kau}}, \bibinfo {author} {\bibfnamefont
  {W.}~\bibnamefont {Choi}}, \bibinfo {author} {\bibfnamefont {C.}~\bibnamefont
  {Chen}}, \bibinfo {author} {\bibfnamefont {Z.}~\bibnamefont {Pedramrazi}},
  \bibinfo {author} {\bibfnamefont {T.}~\bibnamefont {Dumslaff}}, \bibinfo
  {author} {\bibfnamefont {A.}~\bibnamefont {Narita}}, \bibinfo {author}
  {\bibfnamefont {X.}~\bibnamefont {Feng}}, \bibinfo {author} {\bibfnamefont
  {K.}~\bibnamefont {M{\"u}llen}}, \bibinfo {author} {\bibfnamefont
  {F.}~\bibnamefont {Fischer}}, \bibinfo {author} {\bibfnamefont
  {A.}~\bibnamefont {Zettl}}, \bibinfo {author} {\bibfnamefont
  {P.}~\bibnamefont {Ruffieux}}, \bibinfo {author} {\bibfnamefont
  {E.}~\bibnamefont {Yablonovitch}}, \bibinfo {author} {\bibfnamefont
  {M.}~\bibnamefont {Crommie}}, \bibinfo {author} {\bibfnamefont
  {R.}~\bibnamefont {Fasel}}, \ and\ \bibinfo {author} {\bibfnamefont
  {J.}~\bibnamefont {Bokor}},\ }\href@noop {} {\bibfield  {journal} {\bibinfo
  {journal} {Nature Communications}\ }\textbf {\bibinfo {volume} {8}},\
  \bibinfo {pages} {633} (\bibinfo {year} {2017})}\BibitemShut {NoStop}%
\bibitem [{\citenamefont {Jacobse}\ \emph {et~al.}(2017)\citenamefont
  {Jacobse}, \citenamefont {Kimouche}, \citenamefont {Gebraad}, \citenamefont
  {Ervasti}, \citenamefont {Thijssen}, \citenamefont {Liljeroth},\ and\
  \citenamefont {Swart}}]{Jacobse2017}%
  \BibitemOpen
  \bibfield  {author} {\bibinfo {author} {\bibfnamefont {P.~H.}\ \bibnamefont
  {Jacobse}}, \bibinfo {author} {\bibfnamefont {A.}~\bibnamefont {Kimouche}},
  \bibinfo {author} {\bibfnamefont {T.}~\bibnamefont {Gebraad}}, \bibinfo
  {author} {\bibfnamefont {M.~M.}\ \bibnamefont {Ervasti}}, \bibinfo {author}
  {\bibfnamefont {J.~M.}\ \bibnamefont {Thijssen}}, \bibinfo {author}
  {\bibfnamefont {P.}~\bibnamefont {Liljeroth}}, \ and\ \bibinfo {author}
  {\bibfnamefont {I.}~\bibnamefont {Swart}},\ }\href {\doibase
  10.1038/s41467-017-00195-2} {\bibfield  {journal} {\bibinfo  {journal}
  {Nature Communications}\ }\textbf {\bibinfo {volume} {8}},\ \bibinfo {pages}
  {119} (\bibinfo {year} {2017})}\BibitemShut {NoStop}%
\bibitem [{\citenamefont {Chen}\ \emph {et~al.}(2020)\citenamefont {Chen},
  \citenamefont {Narita},\ and\ \citenamefont {M{\"u}llen}}]{Zongping2020}%
  \BibitemOpen
  \bibfield  {author} {\bibinfo {author} {\bibfnamefont {Z.}~\bibnamefont
  {Chen}}, \bibinfo {author} {\bibfnamefont {A.}~\bibnamefont {Narita}}, \ and\
  \bibinfo {author} {\bibfnamefont {K.}~\bibnamefont {M{\"u}llen}},\
  }\href@noop {} {\bibfield  {journal} {\bibinfo  {journal} {Advanced
  Materials}\ }\textbf {\bibinfo {volume} {32}},\ \bibinfo {pages} {2001893}
  (\bibinfo {year} {2020})}\BibitemShut {NoStop}%
\bibitem [{\citenamefont {Cai}\ \emph {et~al.}(2010)\citenamefont {Cai},
  \citenamefont {Ruffieux}, \citenamefont {Jaafar}, \citenamefont {Bieri},
  \citenamefont {Braun}, \citenamefont {Blankenburg}, \citenamefont {Muoth},
  \citenamefont {Seitsonen}, \citenamefont {Saleh}, \citenamefont {Feng},
  \citenamefont {M{\"u}llen},\ and\ \citenamefont {Fasel}}]{Cai2010a}%
  \BibitemOpen
  \bibfield  {author} {\bibinfo {author} {\bibfnamefont {J.}~\bibnamefont
  {Cai}}, \bibinfo {author} {\bibfnamefont {P.}~\bibnamefont {Ruffieux}},
  \bibinfo {author} {\bibfnamefont {R.}~\bibnamefont {Jaafar}}, \bibinfo
  {author} {\bibfnamefont {M.}~\bibnamefont {Bieri}}, \bibinfo {author}
  {\bibfnamefont {T.}~\bibnamefont {Braun}}, \bibinfo {author} {\bibfnamefont
  {S.}~\bibnamefont {Blankenburg}}, \bibinfo {author} {\bibfnamefont
  {M.}~\bibnamefont {Muoth}}, \bibinfo {author} {\bibfnamefont {A.~P.}\
  \bibnamefont {Seitsonen}}, \bibinfo {author} {\bibfnamefont {M.}~\bibnamefont
  {Saleh}}, \bibinfo {author} {\bibfnamefont {X.}~\bibnamefont {Feng}},
  \bibinfo {author} {\bibfnamefont {K.}~\bibnamefont {M{\"u}llen}}, \ and\
  \bibinfo {author} {\bibfnamefont {R.}~\bibnamefont {Fasel}},\ }\href@noop {}
  {\bibfield  {journal} {\bibinfo  {journal} {Nature}\ }\textbf {\bibinfo
  {volume} {466}},\ \bibinfo {pages} {470} (\bibinfo {year}
  {2010})}\BibitemShut {NoStop}%
\bibitem [{\citenamefont {Yano}\ \emph {et~al.}(2020)\citenamefont {Yano},
  \citenamefont {Mitoma}, \citenamefont {Ito},\ and\ \citenamefont
  {Itami}}]{Yano2020}%
  \BibitemOpen
  \bibfield  {author} {\bibinfo {author} {\bibfnamefont {Y.}~\bibnamefont
  {Yano}}, \bibinfo {author} {\bibfnamefont {N.}~\bibnamefont {Mitoma}},
  \bibinfo {author} {\bibfnamefont {H.}~\bibnamefont {Ito}}, \ and\ \bibinfo
  {author} {\bibfnamefont {K.}~\bibnamefont {Itami}},\ }\href@noop {}
  {\bibfield  {journal} {\bibinfo  {journal} {The Journal of Organic
  Chemistry}\ }\textbf {\bibinfo {volume} {85}},\ \bibinfo {pages} {4}
  (\bibinfo {year} {2020})}\BibitemShut {NoStop}%
\bibitem [{\citenamefont {Talirz}\ \emph {et~al.}(2017)\citenamefont {Talirz},
  \citenamefont {S\"ode}, \citenamefont {Dumslaff}, \citenamefont {Wang},
  \citenamefont {Sanchez-Valencia}, \citenamefont {Liu}, \citenamefont
  {Shinde}, \citenamefont {Pignedoli}, \citenamefont {Liang}, \citenamefont
  {Meunier}, \citenamefont {Plumb}, \citenamefont {Shi}, \citenamefont {Feng},
  \citenamefont {Narita}, \citenamefont {M\"ullen}, \citenamefont {Fasel},\
  and\ \citenamefont {Ruffieux}}]{Talirz2017}%
  \BibitemOpen
  \bibfield  {author} {\bibinfo {author} {\bibfnamefont {L.}~\bibnamefont
  {Talirz}}, \bibinfo {author} {\bibfnamefont {H.}~\bibnamefont {S\"ode}},
  \bibinfo {author} {\bibfnamefont {T.}~\bibnamefont {Dumslaff}}, \bibinfo
  {author} {\bibfnamefont {S.}~\bibnamefont {Wang}}, \bibinfo {author}
  {\bibfnamefont {J.~R.}\ \bibnamefont {Sanchez-Valencia}}, \bibinfo {author}
  {\bibfnamefont {J.}~\bibnamefont {Liu}}, \bibinfo {author} {\bibfnamefont
  {P.}~\bibnamefont {Shinde}}, \bibinfo {author} {\bibfnamefont {C.~A.}\
  \bibnamefont {Pignedoli}}, \bibinfo {author} {\bibfnamefont {L.}~\bibnamefont
  {Liang}}, \bibinfo {author} {\bibfnamefont {V.}~\bibnamefont {Meunier}},
  \bibinfo {author} {\bibfnamefont {N.~C.}\ \bibnamefont {Plumb}}, \bibinfo
  {author} {\bibfnamefont {M.}~\bibnamefont {Shi}}, \bibinfo {author}
  {\bibfnamefont {X.}~\bibnamefont {Feng}}, \bibinfo {author} {\bibfnamefont
  {A.}~\bibnamefont {Narita}}, \bibinfo {author} {\bibfnamefont
  {K.}~\bibnamefont {M\"ullen}}, \bibinfo {author} {\bibfnamefont
  {R.}~\bibnamefont {Fasel}}, \ and\ \bibinfo {author} {\bibfnamefont
  {P.}~\bibnamefont {Ruffieux}},\ }\href@noop {} {\bibfield  {journal}
  {\bibinfo  {journal} {ACS Nano}\ }\textbf {\bibinfo {volume} {11}},\ \bibinfo
  {pages} {1380} (\bibinfo {year} {2017})}\BibitemShut {NoStop}%
\bibitem [{\citenamefont {Ruffieux}\ \emph {et~al.}(2016)\citenamefont
  {Ruffieux}, \citenamefont {Wang}, \citenamefont {Yang}, \citenamefont
  {S{\'a}nchez-S{\'a}nchez}, \citenamefont {Liu}, \citenamefont {Dienel},
  \citenamefont {Talirz}, \citenamefont {Shinde}, \citenamefont {Pignedoli},
  \citenamefont {Passerone}, \citenamefont {Dumslaff}, \citenamefont {Feng},
  \citenamefont {M{\"u}llen},\ and\ \citenamefont {Fasel}}]{Ruffieux2016}%
  \BibitemOpen
  \bibfield  {author} {\bibinfo {author} {\bibfnamefont {P.}~\bibnamefont
  {Ruffieux}}, \bibinfo {author} {\bibfnamefont {S.}~\bibnamefont {Wang}},
  \bibinfo {author} {\bibfnamefont {B.}~\bibnamefont {Yang}}, \bibinfo {author}
  {\bibfnamefont {C.}~\bibnamefont {S{\'a}nchez-S{\'a}nchez}}, \bibinfo
  {author} {\bibfnamefont {J.}~\bibnamefont {Liu}}, \bibinfo {author}
  {\bibfnamefont {T.}~\bibnamefont {Dienel}}, \bibinfo {author} {\bibfnamefont
  {L.}~\bibnamefont {Talirz}}, \bibinfo {author} {\bibfnamefont
  {P.}~\bibnamefont {Shinde}}, \bibinfo {author} {\bibfnamefont {C.~A.}\
  \bibnamefont {Pignedoli}}, \bibinfo {author} {\bibfnamefont {D.}~\bibnamefont
  {Passerone}}, \bibinfo {author} {\bibfnamefont {T.}~\bibnamefont {Dumslaff}},
  \bibinfo {author} {\bibfnamefont {X.}~\bibnamefont {Feng}}, \bibinfo {author}
  {\bibfnamefont {K.}~\bibnamefont {M{\"u}llen}}, \ and\ \bibinfo {author}
  {\bibfnamefont {R.}~\bibnamefont {Fasel}},\ }\href@noop {} {\bibfield
  {journal} {\bibinfo  {journal} {Nature}\ }\textbf {\bibinfo {volume} {531}},\
  \bibinfo {pages} {489} (\bibinfo {year} {2016})}\BibitemShut {NoStop}%
\bibitem [{\citenamefont {Li}\ \emph {et~al.}(2021)\citenamefont {Li},
  \citenamefont {Sanz}, \citenamefont {Merino-D\'iez}, \citenamefont
  {Vilas-Varela}, \citenamefont {Garcia-Lekue}, \citenamefont {Corso},
  \citenamefont {de~Oteyza}, \citenamefont {Frederiksen}, \citenamefont
  {Pe{\~{n}}a},\ and\ \citenamefont {Pascual}}]{Li2021}%
  \BibitemOpen
  \bibfield  {author} {\bibinfo {author} {\bibfnamefont {J.}~\bibnamefont
  {Li}}, \bibinfo {author} {\bibfnamefont {S.}~\bibnamefont {Sanz}}, \bibinfo
  {author} {\bibfnamefont {N.}~\bibnamefont {Merino-D\'iez}}, \bibinfo {author}
  {\bibfnamefont {M.}~\bibnamefont {Vilas-Varela}}, \bibinfo {author}
  {\bibfnamefont {A.}~\bibnamefont {Garcia-Lekue}}, \bibinfo {author}
  {\bibfnamefont {M.}~\bibnamefont {Corso}}, \bibinfo {author} {\bibfnamefont
  {D.~G.}\ \bibnamefont {de~Oteyza}}, \bibinfo {author} {\bibfnamefont
  {T.}~\bibnamefont {Frederiksen}}, \bibinfo {author} {\bibfnamefont
  {D.}~\bibnamefont {Pe{\~{n}}a}}, \ and\ \bibinfo {author} {\bibfnamefont
  {J.~I.}\ \bibnamefont {Pascual}},\ }\href@noop {} {\bibfield  {journal}
  {\bibinfo  {journal} {Nature Communications}\ }\textbf {\bibinfo {volume}
  {12}},\ \bibinfo {pages} {5538} (\bibinfo {year} {2021})}\BibitemShut
  {NoStop}%
\bibitem [{\citenamefont {Nguyen}\ \emph {et~al.}(2017)\citenamefont {Nguyen},
  \citenamefont {Tsai}, \citenamefont {Omrani}, \citenamefont {Marangoni},
  \citenamefont {Wu}, \citenamefont {Rizzo}, \citenamefont {Rodgers},
  \citenamefont {Cloke}, \citenamefont {Durr}, \citenamefont {Sakai},
  \citenamefont {Liou}, \citenamefont {Aikawa}, \citenamefont {Chelikowsky},
  \citenamefont {Louie}, \citenamefont {Fischer},\ and\ \citenamefont
  {Crommie}}]{Nguyen2017}%
  \BibitemOpen
  \bibfield  {author} {\bibinfo {author} {\bibfnamefont {G.~D.}\ \bibnamefont
  {Nguyen}}, \bibinfo {author} {\bibfnamefont {H.-Z.}\ \bibnamefont {Tsai}},
  \bibinfo {author} {\bibfnamefont {A.~A.}\ \bibnamefont {Omrani}}, \bibinfo
  {author} {\bibfnamefont {T.}~\bibnamefont {Marangoni}}, \bibinfo {author}
  {\bibfnamefont {M.}~\bibnamefont {Wu}}, \bibinfo {author} {\bibfnamefont
  {D.~J.}\ \bibnamefont {Rizzo}}, \bibinfo {author} {\bibfnamefont {G.~F.}\
  \bibnamefont {Rodgers}}, \bibinfo {author} {\bibfnamefont {R.~R.}\
  \bibnamefont {Cloke}}, \bibinfo {author} {\bibfnamefont {R.~A.}\ \bibnamefont
  {Durr}}, \bibinfo {author} {\bibfnamefont {Y.}~\bibnamefont {Sakai}},
  \bibinfo {author} {\bibfnamefont {F.}~\bibnamefont {Liou}}, \bibinfo {author}
  {\bibfnamefont {A.~S.}\ \bibnamefont {Aikawa}}, \bibinfo {author}
  {\bibfnamefont {J.~R.}\ \bibnamefont {Chelikowsky}}, \bibinfo {author}
  {\bibfnamefont {S.~G.}\ \bibnamefont {Louie}}, \bibinfo {author}
  {\bibfnamefont {F.~R.}\ \bibnamefont {Fischer}}, \ and\ \bibinfo {author}
  {\bibfnamefont {M.~F.}\ \bibnamefont {Crommie}},\ }\href@noop {} {\bibfield
  {journal} {\bibinfo  {journal} {Nature Nanotechnology}\ }\textbf {\bibinfo
  {volume} {12}},\ \bibinfo {pages} {1077} (\bibinfo {year}
  {2017})}\BibitemShut {NoStop}%
\bibitem [{\citenamefont {Yazyev}(2010)}]{Yazyev2010}%
  \BibitemOpen
  \bibfield  {author} {\bibinfo {author} {\bibfnamefont {O.~V.}\ \bibnamefont
  {Yazyev}},\ }\href@noop {} {\bibfield  {journal} {\bibinfo  {journal}
  {Reports on Progress in Physics}\ }\textbf {\bibinfo {volume} {73}},\
  \bibinfo {pages} {056501} (\bibinfo {year} {2010})}\BibitemShut {NoStop}%
\bibitem [{\citenamefont {Fujita}\ \emph {et~al.}(1996)\citenamefont {Fujita},
  \citenamefont {Wakabayashi}, \citenamefont {Nakada},\ and\ \citenamefont
  {Kusakabe}}]{Fujita1996}%
  \BibitemOpen
  \bibfield  {author} {\bibinfo {author} {\bibfnamefont {M.}~\bibnamefont
  {Fujita}}, \bibinfo {author} {\bibfnamefont {K.}~\bibnamefont {Wakabayashi}},
  \bibinfo {author} {\bibfnamefont {K.}~\bibnamefont {Nakada}}, \ and\ \bibinfo
  {author} {\bibfnamefont {K.}~\bibnamefont {Kusakabe}},\ }\href@noop {}
  {\bibfield  {journal} {\bibinfo  {journal} {Journal of the Physical Society
  of Japan}\ }\textbf {\bibinfo {volume} {65}},\ \bibinfo {pages} {1920}
  (\bibinfo {year} {1996})}\BibitemShut {NoStop}%
\bibitem [{\citenamefont {Li}\ \emph {et~al.}(2019)\citenamefont {Li},
  \citenamefont {Sanz}, \citenamefont {Corso}, \citenamefont {Choi},
  \citenamefont {Pe{\~{n}}a}, \citenamefont {Frederiksen},\ and\ \citenamefont
  {Pascual}}]{Li2019}%
  \BibitemOpen
  \bibfield  {author} {\bibinfo {author} {\bibfnamefont {J.}~\bibnamefont
  {Li}}, \bibinfo {author} {\bibfnamefont {S.}~\bibnamefont {Sanz}}, \bibinfo
  {author} {\bibfnamefont {M.}~\bibnamefont {Corso}}, \bibinfo {author}
  {\bibfnamefont {D.~J.}\ \bibnamefont {Choi}}, \bibinfo {author}
  {\bibfnamefont {D.}~\bibnamefont {Pe{\~{n}}a}}, \bibinfo {author}
  {\bibfnamefont {T.}~\bibnamefont {Frederiksen}}, \ and\ \bibinfo {author}
  {\bibfnamefont {J.~I.}\ \bibnamefont {Pascual}},\ }\href@noop {} {\bibfield
  {journal} {\bibinfo  {journal} {Nature Communications}\ }\textbf {\bibinfo
  {volume} {10}},\ \bibinfo {pages} {200} (\bibinfo {year} {2019})}\BibitemShut
  {NoStop}%
\bibitem [{\citenamefont {Magda}\ \emph {et~al.}(2014)\citenamefont {Magda},
  \citenamefont {Jin}, \citenamefont {Hagym{\'a}si}, \citenamefont
  {Vancs{\'o}}, \citenamefont {Osv{\'a}th}, \citenamefont {Nemes-Incze},
  \citenamefont {Hwang}, \citenamefont {Bir{\'o}},\ and\ \citenamefont
  {Tapaszt{\'o}}}]{Magda2014}%
  \BibitemOpen
  \bibfield  {author} {\bibinfo {author} {\bibfnamefont {G.~Z.}\ \bibnamefont
  {Magda}}, \bibinfo {author} {\bibfnamefont {X.}~\bibnamefont {Jin}}, \bibinfo
  {author} {\bibfnamefont {I.}~\bibnamefont {Hagym{\'a}si}}, \bibinfo {author}
  {\bibfnamefont {P.}~\bibnamefont {Vancs{\'o}}}, \bibinfo {author}
  {\bibfnamefont {Z.}~\bibnamefont {Osv{\'a}th}}, \bibinfo {author}
  {\bibfnamefont {P.}~\bibnamefont {Nemes-Incze}}, \bibinfo {author}
  {\bibfnamefont {C.}~\bibnamefont {Hwang}}, \bibinfo {author} {\bibfnamefont
  {L.~P.}\ \bibnamefont {Bir{\'o}}}, \ and\ \bibinfo {author} {\bibfnamefont
  {L.}~\bibnamefont {Tapaszt{\'o}}},\ }\href {\doibase 10.1038/nature13831}
  {\bibfield  {journal} {\bibinfo  {journal} {Nature}\ }\textbf {\bibinfo
  {volume} {514}},\ \bibinfo {pages} {608} (\bibinfo {year}
  {2014})}\BibitemShut {NoStop}%
\bibitem [{\citenamefont {Son}\ \emph {et~al.}(2006{\natexlab{b}})\citenamefont
  {Son}, \citenamefont {Cohen},\ and\ \citenamefont {Louie}}]{Son2006}%
  \BibitemOpen
  \bibfield  {author} {\bibinfo {author} {\bibfnamefont {Y.-W.}\ \bibnamefont
  {Son}}, \bibinfo {author} {\bibfnamefont {M.~L.}\ \bibnamefont {Cohen}}, \
  and\ \bibinfo {author} {\bibfnamefont {S.~G.}\ \bibnamefont {Louie}},\ }\href
  {\doibase 10.1038/nature05180} {\bibfield  {journal} {\bibinfo  {journal}
  {Nature}\ }\textbf {\bibinfo {volume} {444}},\ \bibinfo {pages} {347}
  (\bibinfo {year} {2006}{\natexlab{b}})}\BibitemShut {NoStop}%
\bibitem [{\citenamefont {Kan}\ \emph {et~al.}(2008)\citenamefont {Kan},
  \citenamefont {Li}, \citenamefont {Yang},\ and\ \citenamefont
  {Hou}}]{Kan2008}%
  \BibitemOpen
  \bibfield  {author} {\bibinfo {author} {\bibfnamefont {E.-j.}\ \bibnamefont
  {Kan}}, \bibinfo {author} {\bibfnamefont {Z.}~\bibnamefont {Li}}, \bibinfo
  {author} {\bibfnamefont {J.}~\bibnamefont {Yang}}, \ and\ \bibinfo {author}
  {\bibfnamefont {J.~G.}\ \bibnamefont {Hou}},\ }\href@noop {} {\bibfield
  {journal} {\bibinfo  {journal} {Journal of the American Chemical Society}\
  }\textbf {\bibinfo {volume} {130}},\ \bibinfo {pages} {4224} (\bibinfo {year}
  {2008})}\BibitemShut {NoStop}%
\bibitem [{\citenamefont {Hu}\ \emph {et~al.}(2012)\citenamefont {Hu},
  \citenamefont {Sun},\ and\ \citenamefont {Krasheninnikov}}]{Hu2012}%
  \BibitemOpen
  \bibfield  {author} {\bibinfo {author} {\bibfnamefont {X.}~\bibnamefont
  {Hu}}, \bibinfo {author} {\bibfnamefont {L.}~\bibnamefont {Sun}}, \ and\
  \bibinfo {author} {\bibfnamefont {A.~V.}\ \bibnamefont {Krasheninnikov}},\
  }\href {\doibase 10.1063/1.4731624} {\bibfield  {journal} {\bibinfo
  {journal} {Applied Physics Letters}\ }\textbf {\bibinfo {volume} {100}},\
  \bibinfo {pages} {263115} (\bibinfo {year} {2012})}\BibitemShut {NoStop}%
\bibitem [{\citenamefont {Zhang}\ and\ \citenamefont {Wei}(2017)}]{Zhang2017}%
  \BibitemOpen
  \bibfield  {author} {\bibinfo {author} {\bibfnamefont {D.-B.}\ \bibnamefont
  {Zhang}}\ and\ \bibinfo {author} {\bibfnamefont {S.-H.}\ \bibnamefont
  {Wei}},\ }\href@noop {} {\bibfield  {journal} {\bibinfo  {journal} {npj
  Computational Materials}\ }\textbf {\bibinfo {volume} {3}},\ \bibinfo {pages}
  {32} (\bibinfo {year} {2017})}\BibitemShut {NoStop}%
\bibitem [{\citenamefont {Jung}\ and\ \citenamefont
  {MacDonald}(2010)}]{Jung2010}%
  \BibitemOpen
  \bibfield  {author} {\bibinfo {author} {\bibfnamefont {J.}~\bibnamefont
  {Jung}}\ and\ \bibinfo {author} {\bibfnamefont {A.~H.}\ \bibnamefont
  {MacDonald}},\ }\href {\doibase 10.1103/PhysRevB.81.195408} {\bibfield
  {journal} {\bibinfo  {journal} {Physical Review B}\ }\textbf {\bibinfo
  {volume} {81}},\ \bibinfo {pages} {195408} (\bibinfo {year}
  {2010})}\BibitemShut {NoStop}%
\bibitem [{\citenamefont {Wimmer}\ \emph {et~al.}(2008)\citenamefont {Wimmer},
  \citenamefont {Adagideli}, \citenamefont {Berber}, \citenamefont
  {Tom\'anek},\ and\ \citenamefont {Richter}}]{Wimmer2008}%
  \BibitemOpen
  \bibfield  {author} {\bibinfo {author} {\bibfnamefont {M.}~\bibnamefont
  {Wimmer}}, \bibinfo {author} {\bibfnamefont {I.}~\bibnamefont {Adagideli}},
  \bibinfo {author} {\bibfnamefont {S.}~\bibnamefont {Berber}}, \bibinfo
  {author} {\bibfnamefont {D.}~\bibnamefont {Tom\'anek}}, \ and\ \bibinfo
  {author} {\bibfnamefont {K.}~\bibnamefont {Richter}},\ }\href {\doibase
  10.1103/PhysRevLett.100.177207} {\bibfield  {journal} {\bibinfo  {journal}
  {Physical Review Letters}\ }\textbf {\bibinfo {volume} {100}},\ \bibinfo
  {pages} {177207} (\bibinfo {year} {2008})}\BibitemShut {NoStop}%
\bibitem [{\citenamefont {Pizzochero}\ \emph
  {et~al.}(2021{\natexlab{a}})\citenamefont {Pizzochero}, \citenamefont
  {Barin}, \citenamefont {\v{C}er\c{n}evi\v{c}s}, \citenamefont {Wang},
  \citenamefont {Ruffieux}, \citenamefont {Fasel},\ and\ \citenamefont
  {Yazyev}}]{Pizzochero2021a}%
  \BibitemOpen
  \bibfield  {author} {\bibinfo {author} {\bibfnamefont {M.}~\bibnamefont
  {Pizzochero}}, \bibinfo {author} {\bibfnamefont {G.~B.}\ \bibnamefont
  {Barin}}, \bibinfo {author} {\bibfnamefont {K.}~\bibnamefont
  {\v{C}er\c{n}evi\v{c}s}}, \bibinfo {author} {\bibfnamefont {S.}~\bibnamefont
  {Wang}}, \bibinfo {author} {\bibfnamefont {P.}~\bibnamefont {Ruffieux}},
  \bibinfo {author} {\bibfnamefont {R.}~\bibnamefont {Fasel}}, \ and\ \bibinfo
  {author} {\bibfnamefont {O.~V.}\ \bibnamefont {Yazyev}},\ }\href@noop {}
  {\bibfield  {journal} {\bibinfo  {journal} {The Journal of Physical Chemistry
  Letters}\ }\textbf {\bibinfo {volume} {12}},\ \bibinfo {pages} {4692}
  (\bibinfo {year} {2021}{\natexlab{a}})}\BibitemShut {NoStop}%
\bibitem [{\citenamefont {Avsar}\ \emph {et~al.}(2020)\citenamefont {Avsar},
  \citenamefont {Ochoa}, \citenamefont {Guinea}, \citenamefont {\"Ozyilmaz},
  \citenamefont {van Wees},\ and\ \citenamefont {Vera-Marun}}]{Avsar2020}%
  \BibitemOpen
  \bibfield  {author} {\bibinfo {author} {\bibfnamefont {A.}~\bibnamefont
  {Avsar}}, \bibinfo {author} {\bibfnamefont {H.}~\bibnamefont {Ochoa}},
  \bibinfo {author} {\bibfnamefont {F.}~\bibnamefont {Guinea}}, \bibinfo
  {author} {\bibfnamefont {B.}~\bibnamefont {\"Ozyilmaz}}, \bibinfo {author}
  {\bibfnamefont {B.~J.}\ \bibnamefont {van Wees}}, \ and\ \bibinfo {author}
  {\bibfnamefont {I.~J.}\ \bibnamefont {Vera-Marun}},\ }\href@noop {}
  {\bibfield  {journal} {\bibinfo  {journal} {Reviews of Modern Physics}\
  }\textbf {\bibinfo {volume} {92}},\ \bibinfo {pages} {021003} (\bibinfo
  {year} {2020})}\BibitemShut {NoStop}%
\bibitem [{\citenamefont {Han}\ \emph {et~al.}(2014)\citenamefont {Han},
  \citenamefont {Kawakami}, \citenamefont {Gmitra},\ and\ \citenamefont
  {Fabian}}]{Han2014}%
  \BibitemOpen
  \bibfield  {author} {\bibinfo {author} {\bibfnamefont {W.}~\bibnamefont
  {Han}}, \bibinfo {author} {\bibfnamefont {R.~K.}\ \bibnamefont {Kawakami}},
  \bibinfo {author} {\bibfnamefont {M.}~\bibnamefont {Gmitra}}, \ and\ \bibinfo
  {author} {\bibfnamefont {J.}~\bibnamefont {Fabian}},\ }\href@noop {}
  {\bibfield  {journal} {\bibinfo  {journal} {Nature Nanotechnology}\ }\textbf
  {\bibinfo {volume} {9}},\ \bibinfo {pages} {794} (\bibinfo {year}
  {2014})}\BibitemShut {NoStop}%
\bibitem [{\citenamefont {Yazyev}\ and\ \citenamefont
  {Katsnelson}(2008)}]{Yazyev2008}%
  \BibitemOpen
  \bibfield  {author} {\bibinfo {author} {\bibfnamefont {O.~V.}\ \bibnamefont
  {Yazyev}}\ and\ \bibinfo {author} {\bibfnamefont {M.~I.}\ \bibnamefont
  {Katsnelson}},\ }\href {\doibase 10.1103/PhysRevLett.100.047209} {\bibfield
  {journal} {\bibinfo  {journal} {Physical Review Letters}\ }\textbf {\bibinfo
  {volume} {100}},\ \bibinfo {pages} {047209} (\bibinfo {year}
  {2008})}\BibitemShut {NoStop}%
\bibitem [{\citenamefont {Wang}\ \emph {et~al.}(2009)\citenamefont {Wang},
  \citenamefont {Yazyev}, \citenamefont {Meng},\ and\ \citenamefont
  {Kaxiras}}]{Wang2009}%
  \BibitemOpen
  \bibfield  {author} {\bibinfo {author} {\bibfnamefont {W.~L.}\ \bibnamefont
  {Wang}}, \bibinfo {author} {\bibfnamefont {O.~V.}\ \bibnamefont {Yazyev}},
  \bibinfo {author} {\bibfnamefont {S.}~\bibnamefont {Meng}}, \ and\ \bibinfo
  {author} {\bibfnamefont {E.}~\bibnamefont {Kaxiras}},\ }\href@noop {}
  {\bibfield  {journal} {\bibinfo  {journal} {Physical Review Letters}\
  }\textbf {\bibinfo {volume} {102}},\ \bibinfo {pages} {157201} (\bibinfo
  {year} {2009})}\BibitemShut {NoStop}%
\bibitem [{\citenamefont {Bonfanti}\ \emph {et~al.}(2018)\citenamefont
  {Bonfanti}, \citenamefont {Achilli},\ and\ \citenamefont
  {Martinazzo}}]{Bonfanti2018a}%
  \BibitemOpen
  \bibfield  {author} {\bibinfo {author} {\bibfnamefont {M.}~\bibnamefont
  {Bonfanti}}, \bibinfo {author} {\bibfnamefont {S.}~\bibnamefont {Achilli}}, \
  and\ \bibinfo {author} {\bibfnamefont {R.}~\bibnamefont {Martinazzo}},\
  }\href@noop {} {\bibfield  {journal} {\bibinfo  {journal} {Journal of
  Physics: Condensed Matter}\ }\textbf {\bibinfo {volume} {30}},\ \bibinfo
  {pages} {283002} (\bibinfo {year} {2018})}\BibitemShut {NoStop}%
\bibitem [{\citenamefont {Elias}\ \emph {et~al.}(2009)\citenamefont {Elias},
  \citenamefont {Nair}, \citenamefont {Mohiuddin}, \citenamefont {Morozov},
  \citenamefont {Blake}, \citenamefont {Halsall}, \citenamefont {Ferrari},
  \citenamefont {Boukhvalov}, \citenamefont {Katsnelson}, \citenamefont
  {Geim},\ and\ \citenamefont {Novoselov}}]{Elias2009}%
  \BibitemOpen
  \bibfield  {author} {\bibinfo {author} {\bibfnamefont {D.~C.}\ \bibnamefont
  {Elias}}, \bibinfo {author} {\bibfnamefont {R.~R.}\ \bibnamefont {Nair}},
  \bibinfo {author} {\bibfnamefont {T.~M.~G.}\ \bibnamefont {Mohiuddin}},
  \bibinfo {author} {\bibfnamefont {S.~V.}\ \bibnamefont {Morozov}}, \bibinfo
  {author} {\bibfnamefont {P.}~\bibnamefont {Blake}}, \bibinfo {author}
  {\bibfnamefont {M.~P.}\ \bibnamefont {Halsall}}, \bibinfo {author}
  {\bibfnamefont {A.~C.}\ \bibnamefont {Ferrari}}, \bibinfo {author}
  {\bibfnamefont {D.~W.}\ \bibnamefont {Boukhvalov}}, \bibinfo {author}
  {\bibfnamefont {M.~I.}\ \bibnamefont {Katsnelson}}, \bibinfo {author}
  {\bibfnamefont {A.~K.}\ \bibnamefont {Geim}}, \ and\ \bibinfo {author}
  {\bibfnamefont {K.~S.}\ \bibnamefont {Novoselov}},\ }\href {\doibase
  10.1126/science.1167130} {\bibfield  {journal} {\bibinfo  {journal}
  {Science}\ }\textbf {\bibinfo {volume} {323}},\ \bibinfo {pages} {610}
  (\bibinfo {year} {2009})}\BibitemShut {NoStop}%
\bibitem [{\citenamefont {Balog}\ \emph {et~al.}(2010)\citenamefont {Balog},
  \citenamefont {J{\o}rgensen}, \citenamefont {Nilsson}, \citenamefont
  {Andersen}, \citenamefont {Rienks}, \citenamefont {Bianchi}, \citenamefont
  {Fanetti}, \citenamefont {L\ae{}gsgaard}, \citenamefont {Baraldi},
  \citenamefont {Lizzit}, \citenamefont {Sljivancanin}, \citenamefont
  {Besenbacher}, \citenamefont {Hammer}, \citenamefont {Pedersen},
  \citenamefont {Hofmann},\ and\ \citenamefont {Hornek\ae{}r}}]{Balog2010}%
  \BibitemOpen
  \bibfield  {author} {\bibinfo {author} {\bibfnamefont {R.}~\bibnamefont
  {Balog}}, \bibinfo {author} {\bibfnamefont {B.}~\bibnamefont {J{\o}rgensen}},
  \bibinfo {author} {\bibfnamefont {L.}~\bibnamefont {Nilsson}}, \bibinfo
  {author} {\bibfnamefont {M.}~\bibnamefont {Andersen}}, \bibinfo {author}
  {\bibfnamefont {E.}~\bibnamefont {Rienks}}, \bibinfo {author} {\bibfnamefont
  {M.}~\bibnamefont {Bianchi}}, \bibinfo {author} {\bibfnamefont
  {M.}~\bibnamefont {Fanetti}}, \bibinfo {author} {\bibfnamefont
  {E.}~\bibnamefont {L\ae{}gsgaard}}, \bibinfo {author} {\bibfnamefont
  {A.}~\bibnamefont {Baraldi}}, \bibinfo {author} {\bibfnamefont
  {S.}~\bibnamefont {Lizzit}}, \bibinfo {author} {\bibfnamefont
  {Z.}~\bibnamefont {Sljivancanin}}, \bibinfo {author} {\bibfnamefont
  {F.}~\bibnamefont {Besenbacher}}, \bibinfo {author} {\bibfnamefont
  {B.}~\bibnamefont {Hammer}}, \bibinfo {author} {\bibfnamefont {T.~G.}\
  \bibnamefont {Pedersen}}, \bibinfo {author} {\bibfnamefont {P.}~\bibnamefont
  {Hofmann}}, \ and\ \bibinfo {author} {\bibfnamefont {L.}~\bibnamefont
  {Hornek\ae{}r}},\ }\href {\doibase 10.1038/nmat2710} {\bibfield  {journal}
  {\bibinfo  {journal} {Nature Materials}\ }\textbf {\bibinfo {volume} {9}},\
  \bibinfo {pages} {315} (\bibinfo {year} {2010})}\BibitemShut {NoStop}%
\bibitem [{\citenamefont {Balakrishnan}\ \emph {et~al.}(2013)\citenamefont
  {Balakrishnan}, \citenamefont {Kok Wai~Koon}, \citenamefont {Jaiswal},
  \citenamefont {Castro~Neto},\ and\ \citenamefont
  {\"Ozyilmaz}}]{Balakrishnan2013}%
  \BibitemOpen
  \bibfield  {author} {\bibinfo {author} {\bibfnamefont {J.}~\bibnamefont
  {Balakrishnan}}, \bibinfo {author} {\bibfnamefont {G.}~\bibnamefont {Kok
  Wai~Koon}}, \bibinfo {author} {\bibfnamefont {M.}~\bibnamefont {Jaiswal}},
  \bibinfo {author} {\bibfnamefont {A.~H.}\ \bibnamefont {Castro~Neto}}, \ and\
  \bibinfo {author} {\bibfnamefont {B.}~\bibnamefont {\"Ozyilmaz}},\ }\href
  {\doibase 10.1038/nphys2576} {\bibfield  {journal} {\bibinfo  {journal}
  {Nature Physics}\ }\textbf {\bibinfo {volume} {9}},\ \bibinfo {pages} {284}
  (\bibinfo {year} {2013})}\BibitemShut {NoStop}%
\bibitem [{\citenamefont {Gonz\'alez-Herrero}\ \emph
  {et~al.}(2016)\citenamefont {Gonz\'alez-Herrero}, \citenamefont
  {G\'omez-Rodr\'iguez}, \citenamefont {Mallet}, \citenamefont {Moaied},
  \citenamefont {Palacios}, \citenamefont {Salgado}, \citenamefont {Ugeda},
  \citenamefont {Veuillen}, \citenamefont {Yndurain},\ and\ \citenamefont
  {Brihuega}}]{Gonzales2016}%
  \BibitemOpen
  \bibfield  {author} {\bibinfo {author} {\bibfnamefont {H.}~\bibnamefont
  {Gonz\'alez-Herrero}}, \bibinfo {author} {\bibfnamefont {J.~M.}\ \bibnamefont
  {G\'omez-Rodr\'iguez}}, \bibinfo {author} {\bibfnamefont {P.}~\bibnamefont
  {Mallet}}, \bibinfo {author} {\bibfnamefont {M.}~\bibnamefont {Moaied}},
  \bibinfo {author} {\bibfnamefont {J.~J.}\ \bibnamefont {Palacios}}, \bibinfo
  {author} {\bibfnamefont {C.}~\bibnamefont {Salgado}}, \bibinfo {author}
  {\bibfnamefont {M.~M.}\ \bibnamefont {Ugeda}}, \bibinfo {author}
  {\bibfnamefont {J.-Y.}\ \bibnamefont {Veuillen}}, \bibinfo {author}
  {\bibfnamefont {F.}~\bibnamefont {Yndurain}}, \ and\ \bibinfo {author}
  {\bibfnamefont {I.}~\bibnamefont {Brihuega}},\ }\href@noop {} {\bibfield
  {journal} {\bibinfo  {journal} {Science}\ }\textbf {\bibinfo {volume}
  {352}},\ \bibinfo {pages} {437} (\bibinfo {year} {2016})}\BibitemShut
  {NoStop}%
\bibitem [{\citenamefont {Jiang}\ \emph {et~al.}(2007)\citenamefont {Jiang},
  \citenamefont {Sumpter},\ and\ \citenamefont {Dai}}]{Jiang2007}%
  \BibitemOpen
  \bibfield  {author} {\bibinfo {author} {\bibfnamefont {D.-e.}\ \bibnamefont
  {Jiang}}, \bibinfo {author} {\bibfnamefont {B.~G.}\ \bibnamefont {Sumpter}},
  \ and\ \bibinfo {author} {\bibfnamefont {S.}~\bibnamefont {Dai}},\ }\href
  {\doibase 10.1063/1.2715558} {\bibfield  {journal} {\bibinfo  {journal} {The
  Journal of Chemical Physics}\ }\textbf {\bibinfo {volume} {126}},\ \bibinfo
  {pages} {134701} (\bibinfo {year} {2007})}\BibitemShut {NoStop}%
\bibitem [{\citenamefont {Campisi}\ \emph {et~al.}(2020)\citenamefont
  {Campisi}, \citenamefont {Simonsen}, \citenamefont {Thrower}, \citenamefont
  {Jaganathan}, \citenamefont {Hornek\ae{}r}, \citenamefont {Martinazzo},\ and\
  \citenamefont {Tielens}}]{Campisi2020}%
  \BibitemOpen
  \bibfield  {author} {\bibinfo {author} {\bibfnamefont {D.}~\bibnamefont
  {Campisi}}, \bibinfo {author} {\bibfnamefont {F.~D.~S.}\ \bibnamefont
  {Simonsen}}, \bibinfo {author} {\bibfnamefont {J.~D.}\ \bibnamefont
  {Thrower}}, \bibinfo {author} {\bibfnamefont {R.}~\bibnamefont {Jaganathan}},
  \bibinfo {author} {\bibfnamefont {L.}~\bibnamefont {Hornek\ae{}r}}, \bibinfo
  {author} {\bibfnamefont {R.}~\bibnamefont {Martinazzo}}, \ and\ \bibinfo
  {author} {\bibfnamefont {A.~G. G.~M.}\ \bibnamefont {Tielens}},\ }\href@noop
  {} {\bibfield  {journal} {\bibinfo  {journal} {Physical Chemistry Chemical
  Physics}\ }\textbf {\bibinfo {volume} {22}},\ \bibinfo {pages} {1557}
  (\bibinfo {year} {2020})}\BibitemShut {NoStop}%
\bibitem [{\citenamefont {Perdew}\ \emph {et~al.}(1996)\citenamefont {Perdew},
  \citenamefont {Burke},\ and\ \citenamefont {Ernzerhof}}]{PBE}%
  \BibitemOpen
  \bibfield  {author} {\bibinfo {author} {\bibfnamefont {J.~P.}\ \bibnamefont
  {Perdew}}, \bibinfo {author} {\bibfnamefont {K.}~\bibnamefont {Burke}}, \
  and\ \bibinfo {author} {\bibfnamefont {M.}~\bibnamefont {Ernzerhof}},\
  }\href@noop {} {\bibfield  {journal} {\bibinfo  {journal} {Physical Review
  Letters}\ }\textbf {\bibinfo {volume} {77}},\ \bibinfo {pages} {3865}
  (\bibinfo {year} {1996})}\BibitemShut {NoStop}%
\bibitem [{\citenamefont {Grimme}\ \emph {et~al.}(2010)\citenamefont {Grimme},
  \citenamefont {Antony}, \citenamefont {Ehrlich},\ and\ \citenamefont
  {Krieg}}]{DFTD3}%
  \BibitemOpen
  \bibfield  {author} {\bibinfo {author} {\bibfnamefont {S.}~\bibnamefont
  {Grimme}}, \bibinfo {author} {\bibfnamefont {J.}~\bibnamefont {Antony}},
  \bibinfo {author} {\bibfnamefont {S.}~\bibnamefont {Ehrlich}}, \ and\
  \bibinfo {author} {\bibfnamefont {H.}~\bibnamefont {Krieg}},\ }\href
  {\doibase 10.1063/1.3382344} {\bibfield  {journal} {\bibinfo  {journal} {The
  Journal of Chemical Physics}\ }\textbf {\bibinfo {volume} {132}},\ \bibinfo
  {pages} {154104} (\bibinfo {year} {2010})}\BibitemShut {NoStop}%
\bibitem [{\citenamefont {Bonfanti}\ and\ \citenamefont
  {Martinazzo}(2018)}]{Bonfanti2018b}%
  \BibitemOpen
  \bibfield  {author} {\bibinfo {author} {\bibfnamefont {M.}~\bibnamefont
  {Bonfanti}}\ and\ \bibinfo {author} {\bibfnamefont {R.}~\bibnamefont
  {Martinazzo}},\ }\href@noop {} {\bibfield  {journal} {\bibinfo  {journal}
  {Physical Review B}\ }\textbf {\bibinfo {volume} {97}},\ \bibinfo {pages}
  {117401} (\bibinfo {year} {2018})}\BibitemShut {NoStop}%
\bibitem [{\citenamefont {Hornek\ae{}r}\ \emph {et~al.}(2006)\citenamefont
  {Hornek\ae{}r}, \citenamefont {Rauls}, \citenamefont {Xu}, \citenamefont
  {\v{S}ljivan\v{c}anin}, \citenamefont {Otero}, \citenamefont {Stensgaard},
  \citenamefont {L\ae{}gsgaard}, \citenamefont {Hammer},\ and\ \citenamefont
  {Besenbacher}}]{Hornekaer2006}%
  \BibitemOpen
  \bibfield  {author} {\bibinfo {author} {\bibfnamefont {L.}~\bibnamefont
  {Hornek\ae{}r}}, \bibinfo {author} {\bibfnamefont {E.}~\bibnamefont {Rauls}},
  \bibinfo {author} {\bibfnamefont {W.}~\bibnamefont {Xu}}, \bibinfo {author}
  {\bibfnamefont {{\v{Z}}.}~\bibnamefont {\v{S}ljivan\v{c}anin}}, \bibinfo
  {author} {\bibfnamefont {R.}~\bibnamefont {Otero}}, \bibinfo {author}
  {\bibfnamefont {I.}~\bibnamefont {Stensgaard}}, \bibinfo {author}
  {\bibfnamefont {E.}~\bibnamefont {L\ae{}gsgaard}}, \bibinfo {author}
  {\bibfnamefont {B.}~\bibnamefont {Hammer}}, \ and\ \bibinfo {author}
  {\bibfnamefont {F.}~\bibnamefont {Besenbacher}},\ }\href@noop {} {\bibfield
  {journal} {\bibinfo  {journal} {Physical Review Letters}\ }\textbf {\bibinfo
  {volume} {97}},\ \bibinfo {pages} {186102} (\bibinfo {year}
  {2006})}\BibitemShut {NoStop}%
\bibitem [{\citenamefont {Casolo}\ \emph {et~al.}(2009)\citenamefont {Casolo},
  \citenamefont {L{\o}vvik}, \citenamefont {Martinazzo},\ and\ \citenamefont
  {Tantardini}}]{Casolo2009}%
  \BibitemOpen
  \bibfield  {author} {\bibinfo {author} {\bibfnamefont {S.}~\bibnamefont
  {Casolo}}, \bibinfo {author} {\bibfnamefont {O.~M.}\ \bibnamefont
  {L{\o}vvik}}, \bibinfo {author} {\bibfnamefont {R.}~\bibnamefont
  {Martinazzo}}, \ and\ \bibinfo {author} {\bibfnamefont {G.~F.}\ \bibnamefont
  {Tantardini}},\ }\href {\doibase 10.1063/1.3072333} {\bibfield  {journal}
  {\bibinfo  {journal} {The Journal of Chemical Physics}\ }\textbf {\bibinfo
  {volume} {130}},\ \bibinfo {pages} {054704} (\bibinfo {year}
  {2009})}\BibitemShut {NoStop}%
\bibitem [{\citenamefont {Ghio}\ \emph {et~al.}(1980)\citenamefont {Ghio},
  \citenamefont {Mattera}, \citenamefont {Salvo}, \citenamefont {Tommasini},\
  and\ \citenamefont {Valbusa}}]{Ghio1980}%
  \BibitemOpen
  \bibfield  {author} {\bibinfo {author} {\bibfnamefont {E.}~\bibnamefont
  {Ghio}}, \bibinfo {author} {\bibfnamefont {L.}~\bibnamefont {Mattera}},
  \bibinfo {author} {\bibfnamefont {C.}~\bibnamefont {Salvo}}, \bibinfo
  {author} {\bibfnamefont {F.}~\bibnamefont {Tommasini}}, \ and\ \bibinfo
  {author} {\bibfnamefont {U.}~\bibnamefont {Valbusa}},\ }\href {\doibase
  10.1063/1.439855} {\bibfield  {journal} {\bibinfo  {journal} {The Journal of
  Chemical Physics}\ }\textbf {\bibinfo {volume} {73}},\ \bibinfo {pages} {556}
  (\bibinfo {year} {1980})}\BibitemShut {NoStop}%
\bibitem [{\citenamefont {Bell}(1936)}]{Bell1936}%
  \BibitemOpen
  \bibfield  {author} {\bibinfo {author} {\bibfnamefont {R.~P.}\ \bibnamefont
  {Bell}},\ }\href {\doibase 10.1098/rspa.1936.0060} {\bibfield  {journal}
  {\bibinfo  {journal} {Proceedings of the Royal Society of London. Series A -
  Mathematical and Physical Sciences}\ }\textbf {\bibinfo {volume} {154}},\
  \bibinfo {pages} {414} (\bibinfo {year} {1936})}\BibitemShut {NoStop}%
\bibitem [{\citenamefont {Evans}\ and\ \citenamefont
  {Polanyi}(1936)}]{Evans1936}%
  \BibitemOpen
  \bibfield  {author} {\bibinfo {author} {\bibfnamefont {M.~G.}\ \bibnamefont
  {Evans}}\ and\ \bibinfo {author} {\bibfnamefont {M.}~\bibnamefont
  {Polanyi}},\ }\href {\doibase 10.1039/TF9363201333} {\bibfield  {journal}
  {\bibinfo  {journal} {Transactions of the Faraday Society}\ }\textbf
  {\bibinfo {volume} {32}},\ \bibinfo {pages} {1333} (\bibinfo {year}
  {1936})}\BibitemShut {NoStop}%
\bibitem [{\citenamefont {Bonfanti}\ \emph {et~al.}(2011)\citenamefont
  {Bonfanti}, \citenamefont {Casolo}, \citenamefont {Tantardini}, \citenamefont
  {Ponti},\ and\ \citenamefont {Martinazzo}}]{Bonfanti2011}%
  \BibitemOpen
  \bibfield  {author} {\bibinfo {author} {\bibfnamefont {M.}~\bibnamefont
  {Bonfanti}}, \bibinfo {author} {\bibfnamefont {S.}~\bibnamefont {Casolo}},
  \bibinfo {author} {\bibfnamefont {G.~F.}\ \bibnamefont {Tantardini}},
  \bibinfo {author} {\bibfnamefont {A.}~\bibnamefont {Ponti}}, \ and\ \bibinfo
  {author} {\bibfnamefont {R.}~\bibnamefont {Martinazzo}},\ }\href {\doibase
  10.1063/1.3650693} {\bibfield  {journal} {\bibinfo  {journal} {The Journal of
  Chemical Physics}\ }\textbf {\bibinfo {volume} {135}},\ \bibinfo {pages}
  {164701} (\bibinfo {year} {2011})}\BibitemShut {NoStop}%
\bibitem [{\citenamefont {Pizzochero}\ \emph {et~al.}(2015)\citenamefont
  {Pizzochero}, \citenamefont {Leenaerts}, \citenamefont {Partoens},
  \citenamefont {Martinazzo},\ and\ \citenamefont {Peeters}}]{Pizzochero2015}%
  \BibitemOpen
  \bibfield  {author} {\bibinfo {author} {\bibfnamefont {M.}~\bibnamefont
  {Pizzochero}}, \bibinfo {author} {\bibfnamefont {O.}~\bibnamefont
  {Leenaerts}}, \bibinfo {author} {\bibfnamefont {B.}~\bibnamefont {Partoens}},
  \bibinfo {author} {\bibfnamefont {R.}~\bibnamefont {Martinazzo}}, \ and\
  \bibinfo {author} {\bibfnamefont {F.~M.}\ \bibnamefont {Peeters}},\
  }\href@noop {} {\bibfield  {journal} {\bibinfo  {journal} {Journal of
  Physics: Condensed Matter}\ }\textbf {\bibinfo {volume} {27}},\ \bibinfo
  {pages} {425502} (\bibinfo {year} {2015})}\BibitemShut {NoStop}%
\bibitem [{\citenamefont {Lieb}(1989)}]{Lieb1989}%
  \BibitemOpen
  \bibfield  {author} {\bibinfo {author} {\bibfnamefont {E.~H.}\ \bibnamefont
  {Lieb}},\ }\href {\doibase 10.1103/PhysRevLett.62.1201} {\bibfield  {journal}
  {\bibinfo  {journal} {Physical Review Letters}\ }\textbf {\bibinfo {volume}
  {62}},\ \bibinfo {pages} {1201} (\bibinfo {year} {1989})}\BibitemShut
  {NoStop}%
\bibitem [{\citenamefont {Yazyev}\ and\ \citenamefont
  {Helm}(2007)}]{Yazyev2007}%
  \BibitemOpen
  \bibfield  {author} {\bibinfo {author} {\bibfnamefont {O.~V.}\ \bibnamefont
  {Yazyev}}\ and\ \bibinfo {author} {\bibfnamefont {L.}~\bibnamefont {Helm}},\
  }\href@noop {} {\bibfield  {journal} {\bibinfo  {journal} {Physical Review
  B}\ }\textbf {\bibinfo {volume} {75}},\ \bibinfo {pages} {125408} (\bibinfo
  {year} {2007})}\BibitemShut {NoStop}%
\bibitem [{\citenamefont {Buchs}\ \emph {et~al.}(2007)\citenamefont {Buchs},
  \citenamefont {Krasheninnikov}, \citenamefont {Ruffieux}, \citenamefont
  {Gr\"{o}ning}, \citenamefont {Foster}, \citenamefont {Nieminen},\ and\
  \citenamefont {Gr\"{o}ning}}]{Buchs2007}%
  \BibitemOpen
  \bibfield  {author} {\bibinfo {author} {\bibfnamefont {G.}~\bibnamefont
  {Buchs}}, \bibinfo {author} {\bibfnamefont {A.~V.}\ \bibnamefont
  {Krasheninnikov}}, \bibinfo {author} {\bibfnamefont {P.}~\bibnamefont
  {Ruffieux}}, \bibinfo {author} {\bibfnamefont {P.}~\bibnamefont
  {Gr\"{o}ning}}, \bibinfo {author} {\bibfnamefont {A.~S.}\ \bibnamefont
  {Foster}}, \bibinfo {author} {\bibfnamefont {R.~M.}\ \bibnamefont
  {Nieminen}}, \ and\ \bibinfo {author} {\bibfnamefont {O.}~\bibnamefont
  {Gr\"{o}ning}},\ }\href {\doibase 10.1088/1367-2630/9/8/275} {\bibfield
  {journal} {\bibinfo  {journal} {New Journal of Physics}\ }\textbf {\bibinfo
  {volume} {9}},\ \bibinfo {pages} {275} (\bibinfo {year} {2007})}\BibitemShut
  {NoStop}%
\bibitem [{\citenamefont {Pizzochero}\ \emph
  {et~al.}(2021{\natexlab{b}})\citenamefont {Pizzochero}, \citenamefont
  {\v{C}er\c{n}evi\v{c}s}, \citenamefont {Borin~Barin}, \citenamefont {Wang},
  \citenamefont {Ruffieux}, \citenamefont {Fasel},\ and\ \citenamefont
  {Yazyev}}]{Pizzochero2020}%
  \BibitemOpen
  \bibfield  {author} {\bibinfo {author} {\bibfnamefont {M.}~\bibnamefont
  {Pizzochero}}, \bibinfo {author} {\bibfnamefont {K.}~\bibnamefont
  {\v{C}er\c{n}evi\v{c}s}}, \bibinfo {author} {\bibfnamefont {G.}~\bibnamefont
  {Borin~Barin}}, \bibinfo {author} {\bibfnamefont {S.}~\bibnamefont {Wang}},
  \bibinfo {author} {\bibfnamefont {P.}~\bibnamefont {Ruffieux}}, \bibinfo
  {author} {\bibfnamefont {R.}~\bibnamefont {Fasel}}, \ and\ \bibinfo {author}
  {\bibfnamefont {O.~V.}\ \bibnamefont {Yazyev}},\ }\href@noop {} {\bibfield
  {journal} {\bibinfo  {journal} {2D Materials}\ }\textbf {\bibinfo {volume}
  {8}},\ \bibinfo {pages} {035025} (\bibinfo {year}
  {2021}{\natexlab{b}})}\BibitemShut {NoStop}%
\bibitem [{\citenamefont {Pizzochero}\ and\ \citenamefont
  {Kaxiras}(2021)}]{Pizzochero2021b}%
  \BibitemOpen
  \bibfield  {author} {\bibinfo {author} {\bibfnamefont {M.}~\bibnamefont
  {Pizzochero}}\ and\ \bibinfo {author} {\bibfnamefont {E.}~\bibnamefont
  {Kaxiras}},\ }\href {\doibase 10.1021/acs.jpclett.0c03677} {\bibfield
  {journal} {\bibinfo  {journal} {The Journal of Physical Chemistry Letters}\
  }\textbf {\bibinfo {volume} {12}},\ \bibinfo {pages} {1214} (\bibinfo {year}
  {2021})}\BibitemShut {NoStop}%
\end{thebibliography}%

\end{document}